\documentclass[modern,preprint]{aastex631}

\newcommand\nh{\textit{New Horizons}}
\usepackage{amsmath}
\usepackage{graphicx} 
\submitjournal{ApJS}

\defcitealias{PSTN-056}{PSTN-056}
\defcitealias{PSTN-055}{PSTN-055}
\defcitealias{PSTN-054}{PSTN-054}
 
\def \MIDAS {Michigan Institute for Data Science, University of Michigan, Ann Arbor, MI 48109, USA}
\def \MPhysics {Department of Physics, University of Michigan, Ann Arbor, MI 48109, USA}
\def \MAstro {Department of Astronomy, University of Michigan, Ann Arbor, MI 48109, USA}
\def \CfA {Center for Astrophysics | Harvard \& Smithsonian, 60 Garden Street, Cambridge, MA 02138, USA}
\def \SwRI {Southwest Research Institute, 1301 Walnut St., Suite 400, Boulder, CO 80302, USA}
\def \CCA {Center for Computational Astrophysics, National Astronomical Observatory of Japan, Osawa 2-21-1, Mitaka, Tokyo, 181-8588, Japan}
\def \UOEH {University of Occupational and Environmental Health, 1-1 Iseigaoka, Yahata, Kitakyushu 807-8555, Japan}
\def \PERC {Planetary Exploration Research Center, Chiba Institute of Technology, 2-17-1 Tsudanuma, Narashino, Chiba 275-0016, Japan}
\def \NAAstro {Department of Astronomy and Planetary Science, Northern Arizona University, PO Box 6010, Flagstaff, AZ 86011, USA}
\def \DiRAC {DiRAC Institute and the Department of Astronomy, University of Washington, Seattle, USA}
\def \SUBARU {Subaru Telescope, National Astronomical Observatory of Japan 650 North A`ohoku Place, Hilo, HI 96720, USA}
\def \VAstro {Department of Astronomy, University of Virginia, P.O. Box 400325, Charlottesville, VA 22904-4325, USA}
\def \NRCC {National Research Council of Canada, Herzberg Astronomy and Astrophysics Research Centre, 5071 W. Saanich Rd. Victoria, BC, V9E 2E7, Canada}
\def \UVAstro {Department of Physics and Astronomy, University of Victoria, Elliott Building, 3800 Finnerty Road, Victoria, BC V8P 5C2, Canada}
\def \PSI {Planetary Science Institute, 1700 East Fort Lowell, Suite 106, Tucson, AZ 85719}
\def \PAstro {Department of Physics and Astronomy, University of Pennsylvania, Philadelphia, PA 19104, USA}
\def \Aerotek {Aerotek / Rubin Observatory, Victoria, BC, Canada}
\def \UBC {Department of Physics \& Astronomy, University of British Columbia, 6224 Agricultural Road, Vancouver, BC V6T~1Z1, Canada}
\def \UTINAM {Institut UTINAM UMR6213, CNRS, Universit\'e Bourgogne Franche-Comt\'e, OSU Theta F-25000 Besa\c con, France}
\def \Rubin {Rubin Observatory, 950 N. Cherry Ave., Tucson, AZ 85719, USA}
\def \Aerotek {Aerotek, Suite 150, 4321 Still Creek Drive, Burnaby, BC  V5C 6S7, Canada}

\def \QAstro {Astrophysics Research Centre, School of Mathematics and Physics, Queen's University Belfast, Belfast BT7 1NN, UK}

\def \Canterbury {School of Physical and Chemical Sciences | Te Kura Mat\={u}, University of Canterbury,  Private Bag 4800, Christchurch 8140, New Zealand}

\def \JHUAPL {Johns Hopkins University Applied Physics Laboratory, 11100 Johns Hopkins Road, Laurel, MD 20723, USA}

\begin{document}

\title{An Extremely Deep Rubin Survey to Explore the Extended Kuiper Belt and Identify Objects Observable by {\it New Horizons}}
\date{18 January 2025}

\author[0000-0001-7032-5255]{JJ Kavelaars}
\affiliation{\NRCC}
\affiliation{\UVAstro}
\affiliation{\UBC}

\author[0000-0003-0854-745X]{Marc W. Buie}
\affiliation{\SwRI}

\author[0000-0001-6680-6558]{Wesley C. Fraser}
\affiliation{\NRCC}
\affiliation{\UVAstro}

\author[0000-0002-9179-8323]{Lowell Peltier}
\affiliation{\NRCC}
\affiliation{\UVAstro}

\author[0000-0001-8821-5927]{Susan D. Benecchi}
\affiliation{\PSI}

\author[0000-0003-0333-6055]{Simon B. Porter}
\affiliation{\SwRI}

\author[0000-0002-3323-9304]{Anne J. Verbiscer}
\affiliation{\SwRI}
\affiliation{\VAstro}

\author[0000-0001-6942-2736]{David W. Gerdes}
\affiliation{\MPhysics}
\affiliation{\MAstro}

\author[0000-0003-4827-5049]{Kevin~J.~Napier}
\affiliation{\MPhysics}
\affiliation{\MIDAS}
\affiliation{\CfA}

\author[0000-0001-9505-1131]{Joseph Murtagh}
\affiliation{\QAstro}

\author[0000-0002-0549-9002]{Takashi Ito}
\affiliation{\CCA}

\author[0000-0003-3045-8445]{Kelsi N. Singer}
\affiliation{\SwRI}

\author[0000-0001-5018-7537]{S. Alan Stern}
\affiliation{\SwRI}

\author[0000-0003-4143-4246]{Tsuyoshi Terai}
\affiliation{\SUBARU}

\author[0000-0002-3286-911X]{Fumi Yoshida}
\affiliation{\UOEH}
\affiliation{\PERC}

\author[0000-0003-3257-4490]{Michele T. Bannister}
\affiliation{\Canterbury}

\author[0000-0003-0743-9422]{Pedro H. Bernardinelli}
\altaffiliation{DiRAC Postdoctoral Fellow}
\affiliation{\DiRAC}

\author[0000-0002-8613-8259]{Gary M. Bernstein}
\affiliation{\PAstro}

\author[0000-0001-7335-1715]{Colin Orion  Chandler}
\affiliation{\DiRAC}
\affiliation{\NAAstro}

\author[0000-0002-0283-2260]{Brett Gladman}
\affiliation{\UBC}


\author[0000-0001-5916-0031]{Lynne Jones}
\affiliation{\Aerotek}
\affiliation{\Rubin}

\author[0000-0003-0407-2266]{Jean-Marc Petit}
\affiliation{\UTINAM}

\author[0000-0003-4365-1455]{Megan E. Schwamb}
\affiliation{\QAstro}

\author[0000-0002-4644-0306]{Pontus C. Brandt}
\affiliation{\JHUAPL}

\author[0000-0002-3672-0603]{Joel W. Parker}
\affiliation{\SwRI}

\begin{abstract}

A proposed Vera C. Rubin Observatory Deep Drilling micro-survey of the Kuiper Belt will investigate key properties of the distant solar system. Utilizing 30 hours of Rubin time across six 5-hour visits over one year starting in summer 2026, the survey aims to discover and determine orbits for up to 730 Kuiper Belt Objects (KBOs) to an $r$-magnitude of 27.5. These discoveries will enable precise characterization of the KBO size distribution, which is critical for understanding planetesimal formation.
By aligning the survey field with NASA's {\it New Horizons} spacecraft trajectory, the micro-survey will facilitate discoveries for the mission operating in the Kuiper Belt.  Modeling based on the Outer Solar System Origin Survey (OSSOS) predicts at least 12 distant KBOs observable with the {\it New Horizons} LOng Range Reconnaissance Imager (LORRI) and approximately three objects within 1~au of the spacecraft, allowing higher-resolution observations than Earth-based facilities. LORRI's high solar phase angle monitoring will reveal these objects' surface properties and shapes, potentially identifying contact binaries and orbit-class surface correlations. The survey could identify a KBO suitable for a future spacecraft flyby.
The survey's size, depth, and cadence design will deliver transformative measurements of the Kuiper Belt's size distribution and rotational properties across distance, size, and orbital class. Additionally, the high stellar density in the survey field also offers synergies with transiting exoplanet studies. 
 
\end{abstract}

\section{Introduction}
The Vera C. Rubin Observatory is nearing operations with the recently completed commissioning camera run. At the time of writing, the Legacy Survey of Space and Time (LSST) will soon be underway with Rubin First Light scheduled for summer of 2025\footnote{https://www.lsst.org/about/project-status visited 30-Dec-2024}. 
When operational with LSSTCam, the 8.36-m Simonyi Survey Telescope will image a 9.6 deg$^2$ circular field of view (FOV) with a 30-second cadence, alternating between $u,g,r,i,z,y$ bands.  
This special issue presents several cadence scenarios developed during the planning phases for the LSST, a critical step toward understanding the optimal operations of this multi-year survey.
The currently planned cadence for LSST \citepalias[see][]{PSTN-056} will enable the discovery of an unprecedented number of small solar system bodies, including Trans Neptunian Objects (TNOs), determining precise orbits and physical properties for factors of many more such bodies than are currently known. 
Rubin will genuinely transform our knowledge of the solar system.

As the start the LSST nears, the details of how Rubin Observatory will operate are becoming apparent, and there now exists the possibility that the facility's scheduling budget may accommodate a set of `micro-surveys' (consuming more than 10 but less than 100 hours) \citepalias{PSTN-054, PSTN-055, PSTN-056}.
For example, time gained by moving from doublets of 15-second exposures to single 30-second exposures, one-snap mode, could provide many 100s of hours for micro-surveys.

At the same time, the solar system research community has a second unique opportunity. 
NASA's {\it New Horizons} Mission was specifically designed to characterize bodies in the Kuiper Belt and has made critical unique contributions, including: (1) Understanding the Pluto-Charon system at an unprecedented level \citep[e.g.][]{Stern2019, Stern2015, DESCH2017, WONG2017, KRASNOPOLSKY2020}.
(2) Transforming our understanding of small KBOs \citep{Stern2019} yielding new constraints on chemistry \citep{Grundy2020}, cratering and KBO size distributions \citep{Singer2019, Robbins21}, and processes associated with planetesimal accretion \citep[e.g.][]{Nesvorny2021, Nesvorny2022, Nesvorny2023, Nesvorny2023PSJ, Stern2019, Stern2023, McKinnon2020}. (3) Enabling studies of small KBO properties (shapes, close satellites, surface characteristics) in ways not otherwise achievable \citep{Porter2016, Verbiscer2019, Verbiscer2022}. 

The spacecraft, currently $\sim$61~au from the Sun, continues to be healthy as it moves through the distant Kuiper Belt at a rate of $\sim$3 au/year for another $\sim$15 years. 
Like all missions, {\it New Horizons} has answered many science questions and raised others. 
Although the probability of finding a suitable target is low, the close flyby of a more distant KBO, some 2-3 times further from the Sun than Arrokoth, would be an unprecedented and extraordinary scientific and exploration opportunity. 
Depending on the orbit, these objects could provide {\it in situ} measurement of a surface that has experienced a different thermal history than experienced by either Pluto or Arrokoth, thus providing a revolutionary perspective on the evolution of these otherwise inaccessible bodies. {\it New Horizons} is the only spacecraft currently operating in the Kuiper Belt providing a once in a generation opportunity to observe distant KBOs up close.

Here, we consider the opportunity to use a deep drilling cadence to conduct a Rubin micro-survey to transform our understanding of the Kuiper Belt and find new targets for {\it New Horizons} to observe: the New Horizons Deep Drilling Field (NHDDF). 
This white paper builds on the deep drilling field for solar system science described in \citet{Trilling18} utilizing recent Rubin calibration and operational information, focusing on coordination on the NHDDF, and using recent population models developed by the Outer Solar System Origins Survey \citep[OSSOS][]{Bannister2018}.
The currently planned \href{https://www.lsst.org/scientists/survey-design/ddf}{Rubin Deep Drilling Fields} do not provide any significant possibility for KBO discoveries as they occur at high ecliptic latitude where the sky-density of KBOs is very low.
The field closest to the ecliptic plane, COSMOS, at $8^\circ$ ecliptic latitude, will be blind to the cold classical Kuiper Belt, a population critical to understanding planetesimal formation.    
The proposed observing cadence is to stare at a single LSST pointing for the maximum low-airmass time available in the NHDDF, namely $\sim$5~hr, on 6 nights/visits to secure orbits and basic classification of the detected sample. 
Such a project would deliver both unique science on its own and enable the exploitation of {\it New Horizons'} unique capability to observe from within the Kuiper Belt while providing a chance to find a new distant encounter target for the spacecraft to explore.
Independent of this opportunity, this micro-survey will illuminate fundamental details of this region of space that have never been feasible to investigate and complement the LSST goals. 

\section{Deep Drilling in the Solar System with Rubin}

\begin{figure}
\plotone{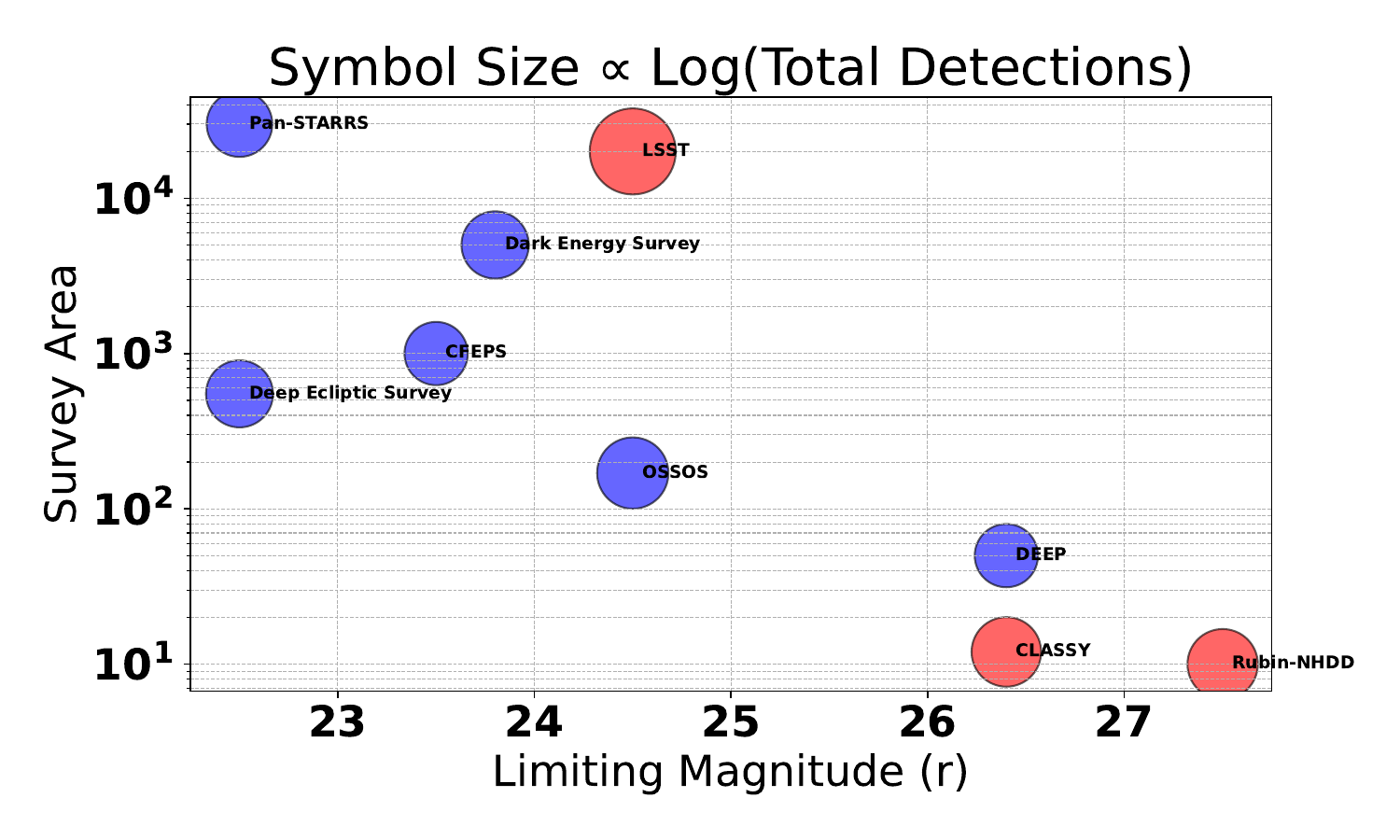}
\caption{\label{fig-surveys}
Coverage and depth from various published (blue) and ongoing or planned (red) TNO surveys.  The proposed project (Rubin-NHDDF) would reach deeper into the KBO population, opening a new component of this phase space.
}
\end{figure}

Wide-area ground-based TNO surveys have systematically reached a depth of $r\simeq26.5$ \citep[see][for recent examples]{Fraser2024PSJ, Yoshida2024, Napier2023}.
These surveys use digital tracking techniques, often combined with machine learning-based image recognition, to `shift-and-stack' multiple short exposures.
In particular, \citet{Fraser2024PSJ} utilize the {\texttt kbmod} \citep{Smotherman2021} routine, which was specifically developed to operate on images processed through the Rubin LSST Science Pipeline.
\citet{Fraser2024PSJ} conducted their search in the sky location that is coincident with the {\it New Horizons} trajectory.
They have demonstrated the ability to achieve near photon-noise-limit depths on full stacks of digitally tracked images.
As outlined in Section~\ref{sec:strategy}, a single field Rubin micro-survey using modern digital tracking should achieve a depth of $r \sim 27.5$, the deepest ground-based KBO survey to date, discovering over 700 new KBOs. 
New search techniques that enable multi-night stacking (such as reported by Napier at TNO 2024\footnote{https://tno2024.org/relation/abstract/46}) could extend the limit of detection even further, perhaps as deep as $r\simeq28.2$.
Even at slightly shallower depths, Rubin will significantly benefit {\it New Horizons}, studies of the Kuiper Belt, and studies of other solar system bodies. 
Figure~\ref{fig-surveys} presents a view of the currently available, ongoing, or planned KBO surveys; this Rubin micro-survey would probe new depths and provide the significant populations needed to enable new science.

\subsection{Kuiper Belt Science}

\begin{figure}
\plottwo{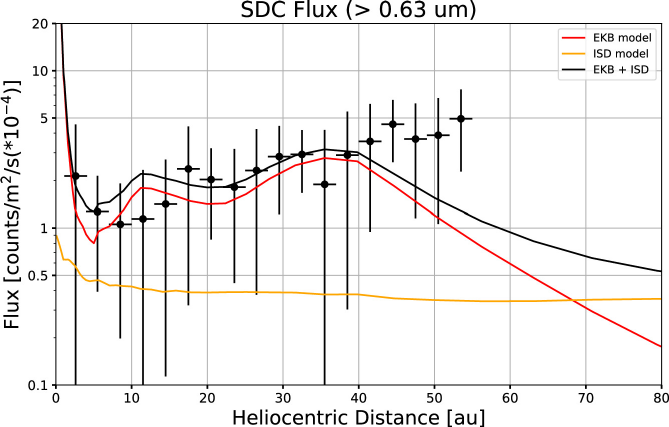}{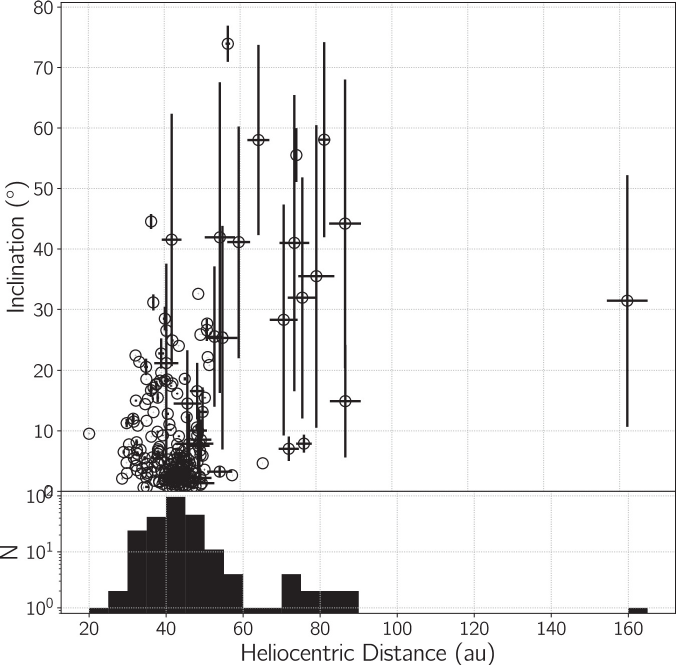}
\caption{\label{fig-KBOsPlot}(left) Figure 3 from \citet[][used with permission]{Doner2024}, SDC flux estimates for particles with a radius greater than 0.63 $\micron $ from 1 to 55 heliocentric astronomical units (au). Each point averages the flux measured by each film across each 3~au traversed by the {\it New Horizons} spacecraft. (right) Figure 1 from \citet[][used with permission]{Fraser2024LPI}, showing the inclination and distance of the {\it New Horizons} discoveries from Subaru Telescope Hyper Suprime-Cam (HSC) observations 2020-2023. A histogram of the heliocentric distances is shown in the bottom panel, with a bin width of 5~au chosen to be similar to the typical uncertainty of the objects with best-fit heliocentric distance $R > 70$ au. These data suggest that many KBOs are located ahead of {\it New Horizons} and may be accessible to it for observation.}

\end{figure}

\begin{figure}
\plotone{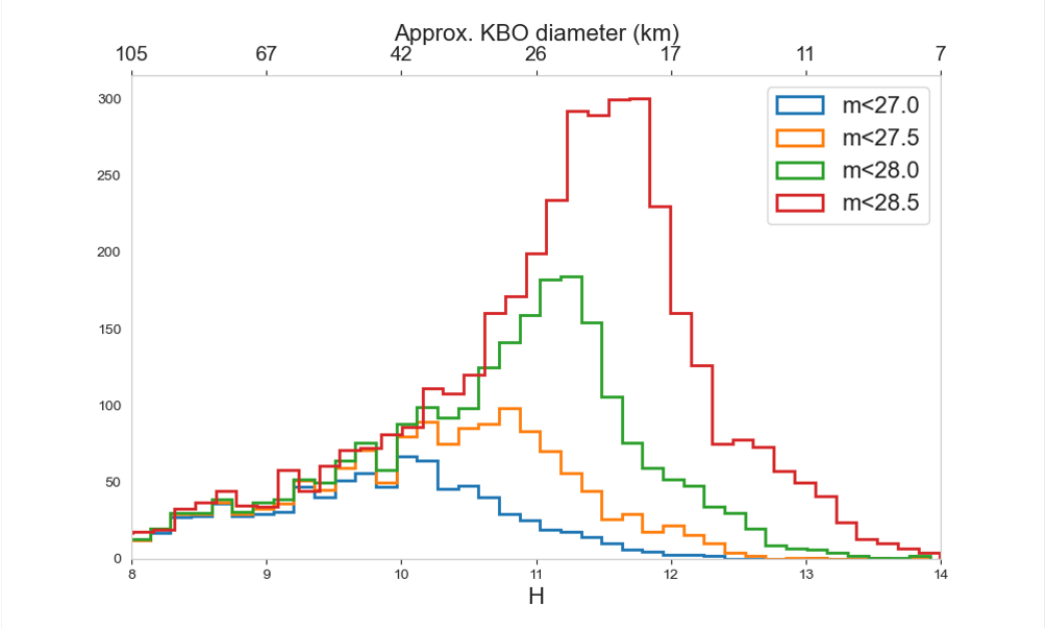}
\figcaption{\label{fig-Hmag}
H magnitude distribution for KBOs discoverable at the {\it New Horizons} search location as a function of survey depth. 
The input model is the OSSOS$++$ model and does not include the putative distant population; see text for details. 
The corresponding size of these objects, assuming an albedo of $p_V=0.10$, is shown along the top axis. The y-axis is the number of objects discovered at each H magnitude/size based on survey magnitude depth, denoted by color shown in the inset key. 
}
\end{figure}

\begin{figure}
\plotone{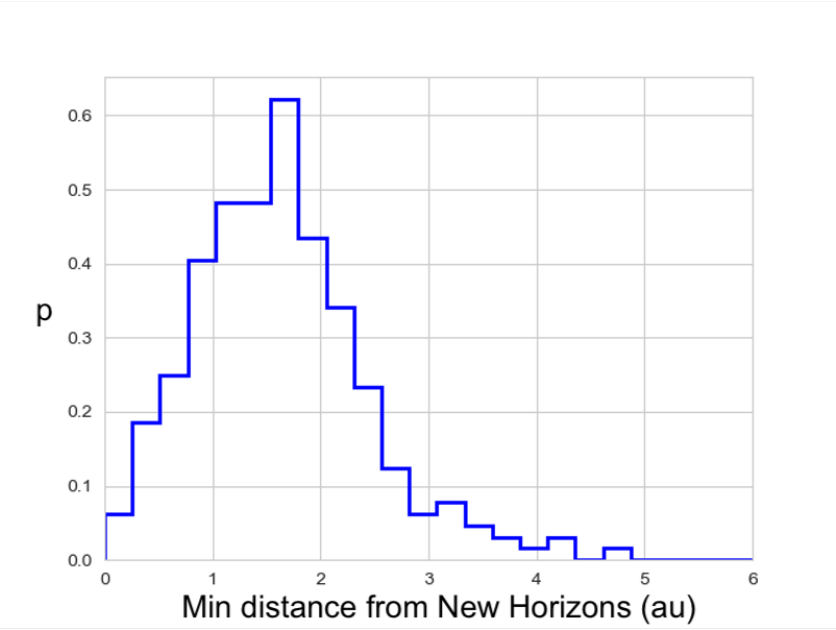}
\figcaption{\label{fig-HmagLORRI}
The probability density (p) vs.\ discovery distance from the {\it New Horizons} spacecraft for Rubin discoveries with projected LORRI magnitudes of $V<20.5$ and currently beyond 70~au. Approximately 40\% of objects discovered beyond 70~au will come within 1~au of the spacecraft.}
\end{figure}

The Rubin Observatory LSST Camera's wide field of view, combined with expected deep drilling depths, enables the characterization of a deep, extended Kuiper Belt in detail for the first time. Current ground-based detections of distant KBOs, see Figure \ref{fig-KBOsPlot} \citep{Fraser2024LPI, Yoshida2024}, as well as measurements by the {\it New Horizons} Venetia Burney Student Dust Counter \citep[SDC,][]{Doner2024}, give evidence for a potentially sizable component to the disk further out. 
The $>$700 KBOs (Table \ref{tab:timetable}) that this proposed 30-hour observation program, 5 hours per visit over six visits spanning two oppositions, of the NHDDF will detect include a significant fraction of small objects with $H<11$ (or smaller, dependent on cumulative observing time and sky conditions, see Section~\ref{sec:strategy} for details).
These small objects probe a region of the size distribution, see Figure $\ref{fig-Hmag}$, which has been historically limited to space-based telescopes, which come at an exceedingly high observing cost despite the woefully tiny fields of view.  
Figure $\ref{fig-HmagLORRI}$ shows the probability function vs.\ discovery distance from {\it New Horizons} for Rubin discoveries with projected magnitudes of $V<20.5$ in the spacecraft's LOng Range Reconnaissance Imager (LORRI) \citep{2008SSRv..140..189C,2020PASP..132c5003W}. A \textit{New Horizons} close flyby target would need to be discovered within about 0.03~au of the spacecraft's current trajectory. Determining the precise likelihood of a flyby is contingent on a more secure determination of the population of the extended Kuiper Belt.

\subsection{{\it New Horizons} Science}
At 61~au, {\it New Horizons} is uniquely positioned to explore KBOs in the outer solar system.
Owing to the size of the Earth's orbit, objects beyond Neptune are only visible at solar phase angles (Sun-KBO-Observer) $<2^\circ$ from Earth-based observatories.
Observations at larger phase angles are only accessible to spacecraft far from Earth, and {\it New Horizons} is the only spacecraft planned to fly between 61 and 106~au for decades.
{\it New Horizons} LORRI has already observed about three dozen KBOs at high phase angles, characterizing their surface scattering properties, shapes, rotation poles, and periods \citep{Porter2016, Verbiscer2019, Verbiscer2022}.
By applying photometric models to disk-integrated solar phase curves of KBOs, {\it New Horizons} has already found correlations between phase function and surface composition \citep{Verbiscer2022}. 
KBOs with highly volatile ices on their surfaces have shallower phase curves than those without volatiles, like Arrokoth. 
These high phase angle observations provide insight into object composition and, by inference, its formation location.
Six of these objects were observe when closer than 1~au to the space craft, one was found to a tight binary on a presumably circular orbit, another a contact binary and a third very likely a binary \citep{2022PSJ.....3...46W}.
Rotation light curves acquired at multiple high phase angles have also revealed a high fraction of contact binaries \citep{Porter2024}, suggesting that more than 50\% of the small KBO population are in multiple or contact systems. The closeness of these binary pairs and pole projections of the contact systems required the resolution and/or view geometry of LORRI to be revealed.
Since {\it New Horizons} has left the densely populated classical Kuiper Belt, the KBOs that this Rubin micro-survey would find for study with LORRI are members of more distant populations, some with apoapses that extend beyond the heliopause.
Modelling based on the known Kuiper Belt indicate that LORRI would be able to view the $\sim$3 targets within 1~au, providing high spatial resolution for tight satellite discover and 9 additional targets at a longer range. 

\subsection{The Distant Kuiper Belt}

At the time of this writing, numerous ultra-deep surveys are ongoing. The main goals of these surveys are to measure the size distribution of Kuiper Belt populations down to much smaller sizes and with significantly higher fidelity than previously done (e.g., \citet{Smotherman2024}, see \citet{Lawler24} for a recent summary). 
One of the surprising results coming from \emph{some} of these surveys is the potential discovery of a heretofore unrecognized massive population of distant bodies. 
Observations presented by \citet{Fraser2024PSJ} imply distant regions which were previously thought to be relatively empty, have a factor of $\sim4$ more objects at a distance $r\gtrsim70$~au than detected in previous surveys \citep{Bannister2018}. 
Such a population would rival the most massive subpopulations of more proximate KBO populations, such as the population of 3:2 mean-motion resonators, and would represent a significant increase in the bounds of what was previously considered the outer bound of the known Kuiper Belt ($\sim$55~au;  referred to as the Kuiper cliff). Such objects could even be the start of the Kuiper wall representing a rebound in the primordial surface density of the Sun's proto-planetary disk \citep{Chiang1999}.
A putative Kuiper wall can have escaped detection under the assumption that such a distant population may have formed with a truncated size distribution that, as did the cold Classical Kuiper belt, and is devoid of objects with $H_r \lesssim 5$ \citep{Gladman2022, Peltier2022}.
Wide area surveys would need to have reached depths of  $r<24.5$ to have found the very largest objects, of which there would only be a very few in such a population. 

Fundamentally, this Rubin micro-survey will elucidate the distant regions beyond the main Kuiper Belt in ways otherwise currently unobtainable \citep{Kavelaars2020}. 
At 70~au objects that with $r=27$ have $H_r \sim 8.5$
Suppose the distant bodies' size frequency distribution (SFD) follows the slope typical for KBOs in a similar size range ($N(H) \propto 10^{\alpha H}, \alpha\sim0.6$). 
In that case, the sky density of these distant objects at magnitude $r\simeq27.5$ will be roughly 4 times higher than at the current state-of-the-art Subaru depths ($r<26.5$). 
As such, considering the Rubin areal coverage and depth, we would expect $\sim$80 newly-detected objects at and beyond 70~au (\citet{Fraser2024PSJ} reported 11 in 5 deg$^2$ compared to Rubin's 11deg$^2$ FOV) 
{\bf This Rubin micro-survey would confirm this distant population and enable its first direct characterization. 
Providing the tracking observations needed to ensure a high level of veracity to the detections of these distant bodies and confirm the previously reported over-density, compared to the OSSOS$++$ model \citep{Petit2023DPS}\footnote{The OSSOS$++$ Kuiper Belt model assembles the classical Kuiper Belt, resonant, scattering and detached populations described in a series of papers from the Outer Solar System Origin Survey group \citet{Bannister2018, Lawler2018, Kavelaars2021, Crompvoets2022, Petit2023, beaudoin2023}.}}. 

If these distant KBOs' origins are with the dynamically excited KBOs, then one would expect them to have moderate to high inclinations and eccentricities, and their orbits would be coupled to Neptune. 
Objects in resonance with Neptune come to their apocenters in the area search by \citet{Fraser2024PSJ} and could be an explanation for their detections. 
If these distant objects formed from a cold disk and were pushed out during smooth migration, they would be found with low inclinations and moderate eccentricities, still dynamically coupled to Neptune, or, if they are from a second Kuiper Belt, they would be entirely uncoupled from Neptune. Current data poorly constrain the orbital distribution of the distant objects reported by \citet{Fraser2024PSJ}.  
In these circular orbit cases one might then expected a truncated size distribution has hidden them from the view of wide area surveys with limits of $<24.5$.

\subsection{Size Distribution, KBO Formation \& Cratering} 
Through the brightness range $r<26$, where the sample of KBOs has been measured robustly, we see a size distribution that is compatible with expectations from modern planetesimal growth models, such as the streaming instability \citep{Kavelaars2021, Petit2023, Napier2023}. 
Available surveys demonstrate that the number density distribution of objects for object smaller than $\sim$400~km diameter through the observable range results in a uniform total mass per unit size. 
As such, observations of objects in this size range provide limited leverage on proto-planetary disk models. 
Only at the faint end of what is reachable from current telescope offerings does the distribution exhibit deviations away from equal mass per size bin \citep{Napier2023}. 
These deviations are critical in elucidating the formative processes outer Solar System objects have experienced, but available measurements come with extremely poor fidelity simply due to the low numbers of small KBOs detected to date; at the time of this writing, only $\sim$10 KBOs fainter than $r=26.5$ have ever been detected with robust orbit determination \citep{Bernstein2004, Smotherman2024}. 
Simulations of macroscopic bodies formed via streaming instability predict that below a specific size, objects are no longer the primary accretion products of pebble cloud collapse but rather are the unaccreted detritus that is ejected during collapse \citep[e.g.,][]{Robinson2020, Polak2023}. 
To probe this process requires reaching sizes below those produced by pebble cloud collapse providing evidence for how collisionally evolved this small size material is, how high the collisional interactions during accretion were, and what the accretion efficiency (mass of detritus to macroscopic bodies) is. 
All of these are fundamental to understanding planetesimal growth and currently remain out of reach, at least until this micro-survey is executed.

There are no known plans to conduct surveys that go both broad and deep enough that they can directly connect the large ($D>$ 200-km) object slope and small ($D<$ 20-km) object slopes to provide a robust measure of the preferred size scale of the Kuiper Belt's planetesimal population.
This Rubin micro-survey will provide a sample of KBOs that vastly improve upon the known faint sample while simultaneously directly connecting to the bright object distribution ($\sim$10 objects with $r<23.3$ will be in the field of view). A survey to $r<27.5$ will provide a sample of $\lesssim 730$ discoveries, with the majority being fainter/smaller than where the fidelity of the known sample crumbles. The discovered sample will increase by a factor of $\sim$2$\times$ for every $\sim0.5$ magnitude increase in depth limit. 

\subsection{Rotational Light Curves}
The planned cadence will also provide an exceptional measure of the rotational properties of the KBOs, another indicator of the planetesimal formation process..
Exposures for each of the six visits will be 30 seconds, resulting in 540 exposures per 4.5 hours of integration (taking 5 hours with overheads) at each visit.  
Combining data across and within visits will enable the determination of (partial) rotational light curves \citep[see][for example]{Strauss24} for those sources brighter than about the 10$\sigma$ limit in the stack (i.e., $r<26$).  
This will provide some shape information for $\sim$100 KBOs. 
The NHDDF field is included in the planned coverage of LSST. Based on the recent operational simulations (one\_snap\_v3\.6\_10yrs.db\footnote{https://usdf-maf.slac.stanford.edu/allMetricResults?runId=16}), this field will be visited $\sim$100 times per filter throughout the survey.  
Based on simulations using one\_snap\_v3\.6\_10yrs.db, the OSSOS$++$ model, and \texttt{Sorcha} (see Section~\ref{sec:strategy} for details), we expect the LSST project will provide an orbital classification for the $\sim30$ KBOs within the NHDDF.  
For this sample of objects, the deep sampling provided by the NHDDF fields will enable detailed studies of the rotational state of the objects. This information can be used to explore further the relationship between object shape and size, an indicator of collisional evolution processes in the outer solar system \citep{Benechhi13}.

\subsection{Occultation Opportunities}
Near galactic coordinates $l=17.5$, $b=-14.8$, the NHDDF is near the galactic center and has a high stellar density (see Figure~\ref{fig-NH_Placement}).  Although challenging, previous searches \citep[e.g.][]{Fraser2024PSJ} have demonstrated the feasibility of difference imaging in this part of the sky, achieving near photon-noise-limit detections using the LSST Science Pipeline's differencing engine.
At the same time, the high stellar density provides a high likelihood that many newly discovered objects will have their physical sizes measured via predicted stellar occultations. Among the sample of discovered objects that have occultations, some will be drawn from the $\sim 100$ KBOs we expect to have determined rotational light curves of, providing the opportunity to link these objects' light-curve shapes and occultation shapes. Existing occultation facilities \cite[e.g.][]{RECON16} and teams will pursue many of these opportunities. Occultation-based shaped measurements will provide a solid link between the physical shapes of these KBOs and their rotational light curves, further enhancing our knowledge of the formation and evolution processes at work in the Kuiper Belt.

\section{Solar System Deep Drilling Observing Strategy\label{sec:strategy}}

\subsection{Cadence}
This survey aims to detect the faintest moving sources within a micro-survey budget while obtaining sufficient visits to enable moderately precise orbits.
Sky motion is a few arcseconds per hour for distant solar system bodies (those beyond 30~au). 
For such distant bodies, measurements at four epochs in a single observing season, two visits in each of two different lunations, are sufficient to achieve the ephemeris precision needed to enable linking in the following year.  
In the following year, one must obtain measurements at two epochs to secure the recovery of the object.  
Given the field location, the opposition observations (two visits) should occur in July, while the two tracking visits would be best scheduled in September, and the recovery one year later should occur over two visits separated by a few nights and taken between May and August of the following year. 
With these two epochs of observation, one can compute accurate distances and orbital inclinations and have some security in the determination of the semi-major axis of the orbit [see, for example, Figure 10, \citet{Bannister2016}].
This information is sufficient for determining the size distribution, in particular allowing the separation of objects into the two main KBO orbital groups of `cold' and `excited' orbits.
Those objects found beyond 70~au, of which the OSSOS$++$ model predicts as many as $\sim40$, can be pursued using large aperture and space-based facilities with smaller fields of view but greater sensitivity (e.g., Hubble Space Telescope, JWST, Gemini, Magellan) 
Visits of 5-hour integrations would allow for the six needed visits within the expected 30-hour limit for micro-surveys. 

\subsection{Depth}
Given the interplay between object reflectivity and solar flux and the desire to achieve the faintest (small and or distant) detections possible within the time constraints, the survey observations will be obtained in a single Rubin filter, $r$.
A depth of $r\sim 27.5$ can be achieved in each of the six visits to a single Rubin field of view with a 30-hour investment of observing time.
For the moving source discovery and tracking cadence proposed here, we limit our total exposures to be within a single night and when the field is above an airmass of 1.5, providing $\sim$4.5~hours of exposure time per night both when the field is at opposition and 8 weeks later when the field would be re-observed to enable recovery observations of sources detected during opposition.
Utilizing the latest through-put information expressed in \texttt{rubin\_scheduler.utils.m5\_scale}\footnote{https://rubin-scheduler.lsst.io/}, sky brightness from the ESO Sky Model\footnote{https://www.eso.org/observing/etc/bin/gen/form?INS.MODE=swspectr+INS.NAME=SKYCALC}, an average airmass of 1.3 and 540 30-second exposures predicts a stack depth of $m_r \sim 27.5$.
A solar system object survey to this depth over this area of sky with built-in tracking will be unprecedented.

\subsection{Field Selection}
To enable detection of sources that can be observed from {\it New Horizons} requires observing the field near RA=289.4, DEC=$-20.2$ (Figure \ref{fig-NH_Placement}). 
In addition, other deep survey's have not 
This field is defined as the search region for objects that are at least 66~au from the Sun between January 1, 2027 and 2040, and are within 7 degrees on the sky of {\it New Horizons'} location as seen from the Earth in July 2026 (see Figure~\ref{fig-RubinField}) and will pass within 1~au of {\it New Horizons} before 2040 and have $r <28.5$ in July 2026. 
Model orbits are drawn from the OSSOS$++$ Kuiper Belt model. 

Previous surveys at this longitude have revealed an over density of sources while deep surveys at other longitudes \citep[e.g.][]{Napier2023} have not reported an over density of distant objects. This may indicate that the \citet{Fraser2024PSJ} detections are Neptune resonant objects coming to apocenter at this longitude, the Rubin NHDDF survey is at this longitude and will reveal if this is the case.

Further to providing targets for LORRI this field selection the high stellar density in the field enables the occultation science case as well as the possibility of conducting a parallel surveys for variable stars and transiting planets near the disk of the Milky Way.  

\subsection{Expected Detections}
We used the 1.0b release of the \texttt{Sorcha} (Merritt {\it et al., submitted}, Holman {\it et al., submitted}, see Software section) package to simulate the impact of our cadence, total integration time and field choice on the number of KBOs detected. 
\texttt{Sorcha} computes the ephemeris of the objects given an orbit input file and determines if the object lands on the footprint of LSSTCam given a database of pointings.  
For sources that are on the footprint, \texttt{Sorcha} determines the flux, including sky and shot noise, Rubin would measure in the $r$ filter using an assigned H value, the ephemeris, and the phase angle of the observation (we used \texttt{Sorcha's} \texttt{HG} phase model with a slope of 0.15). 
In our use, we modeled the NHDDF observations as a series of visits to a single Rubin pointing where each visit consisted of a single 30s exposure whose 5-$\sigma$ limit was varied between $27$ and 28.5 in $r$ using \texttt{Sorcha's} fading function to express the shape of the detection efficiency curve setting the width to 0.15 and peak detection efficiency to 0.85 (to account for some area loss due to nearby bright stars).  
A short exposure time ensured no trailing loss (which \texttt{Sorcha} can simulate).  
The six visits were spaced with two near opposition, two about 8 weeks later and two about one year later. 
Each visit was assigned an image quality of 0.8 arcseconds, typical of expectations for Rubin.
The \texttt{Sorcha} tracklet linking step was not used. Instead, we collated the object and field ID values to determine which model objects were detected in each visit. 
We ran 20 realizations of the OSSOS$++$ model for each flux limit.
Based on these simulations (see Table~\ref{tab:timetable}), we find the NHDDF would result in the detection of 780$\pm{50}$ KBOs, and the offset strategy, trailing the field at the mean motion of all sources beyond 50~au, results in $\sim$94\% of all KBOs in the model being recovered in the second opposition.

Based on these simulations, we expect the survey to discover $\sim12$ KBOs observable by LORRI of which $\sim$3 will approach within 1~au of {\it New Horizons}.
Table \ref{tab:timetable} shows the number of KBOs projected to be within 1~au of the spacecraft for different survey depths. Estimates provided here are based on the OSSOS$++$ model and do not include the putative distant population reported in \citep{Fraser2024PSJ}, including that population would double or quadruple the distant detections. 
A search to $m=28.5$ would require a total of 200 hours. Nevertheless, even a micro-survey of 30~hours on the {\it New Horizons} field proposed here yields a significant possibility for objects accessible for observations from the spacecraft. 
If the enhanced population reported in \citep{Fraser2024PSJ}, which we have not used in our detection estimate here, proves to be correct, the number of targets seen within 1~au will be a factor of $\sim4$ higher. 
While the probability of one of those targets being a flyby candidate is low, it is not zero.

\subsection{Scheduling}

As \nh\ moves farther through the belt the opportunity for observations and an encounter diminish, making discovery and followup a high urgency. For all objects that could potentially be targeted by {\it New Horizons} or that are sufficiently scientifically interesting, we will propose follow-up observations with JWST and/or HST to provide additional astrometry to refine their orbits and measure their broad-band colors.
In 2020, the {\it New Horizons} project demonstrated this process with KBO targets discovered by Subaru in June-July, recovered by Subaru in August-September, recovered with HST in November, and then observed by {\it New Horizons} in December.
A similar timeline for objects discovered {\bf in the July-August 2025 or 2026 oppositions} by a Rubin micro-survey would be possible, with rapid follow-up when the field passes through the HST field of regard, starting $\sim$45 days after opposition and lasting $\sim$50 days enabling LORRI observations as early as the spring of 2026 or 2027 (observations with Rubin in summer 2025 enable New Horizons followup in 2026).

\begin{figure}
\plotone{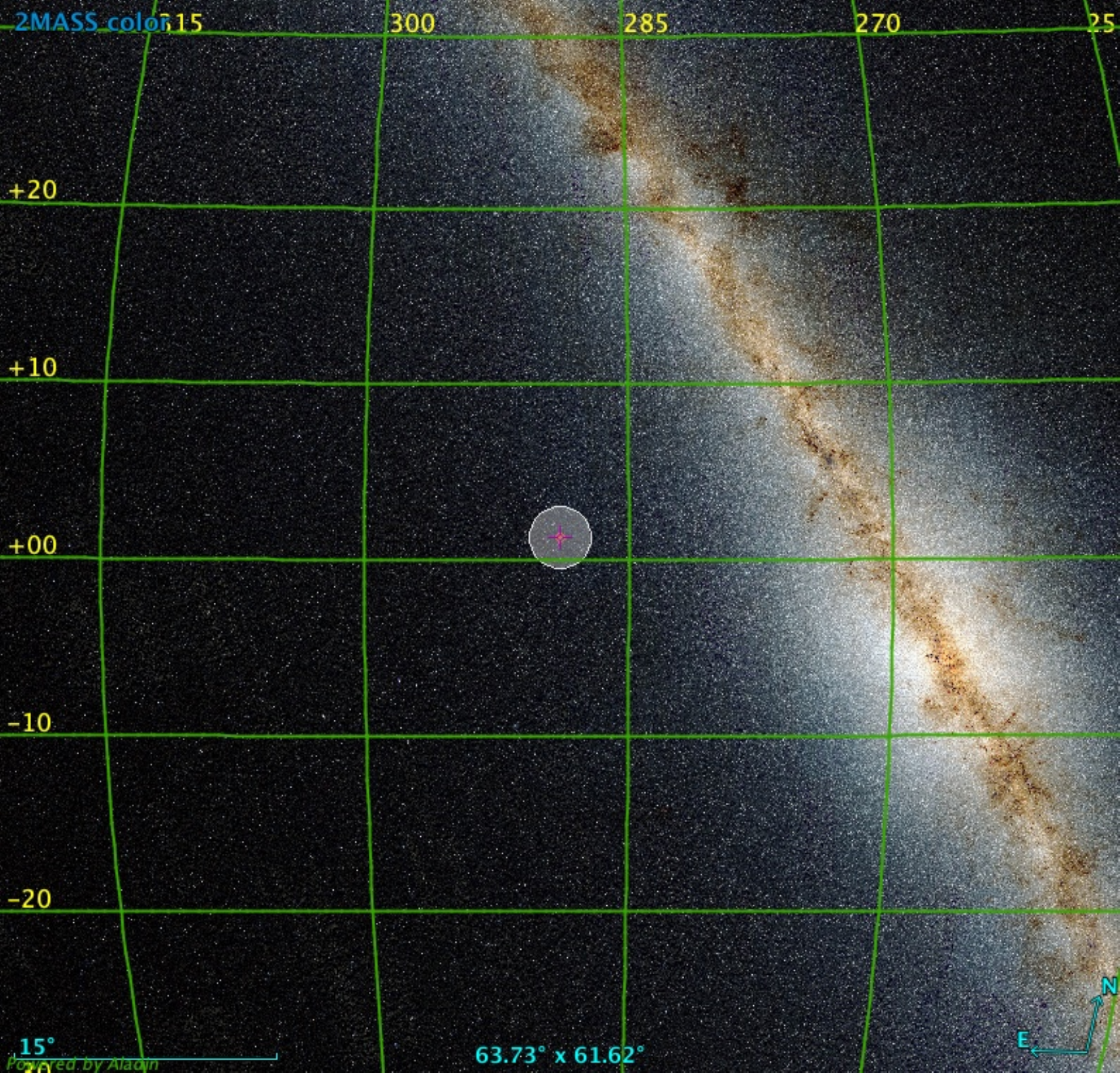}
\figcaption{\label{fig-NH_Placement}
The opaque circle indicates the NHDDF relative to the galactic plane (image from 2MASS \cite{2MASS}) displayed using the Aladin Sky Atlas \citep{ALADIN}. The coordinate grid is latitude by longitude in ecliptic degrees.
}
\end{figure}

\begin{figure}
\plotone{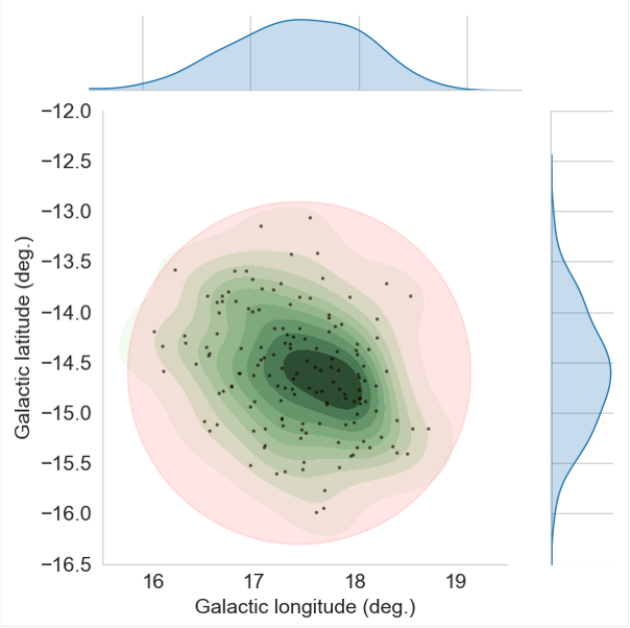}
\caption{\label{fig-RubinField}
{\it New Horizons} search field with the Rubin footprint (pink circle). This plot shows the current sky locations of model objects (black points and sky density contours) that pass within 1~au of {\it New Horizons} between 2027-2040 and have $r<28.5$ in July 2026. The model is based on the OSSOS$++$ sample \citep{Petit2023DPS} and is a 20$\times$ oversampling of the excepted total population of the Kuiper Belt.}
\end{figure}

\begin{deluxetable*}{rrrrr}
\caption{Expected KBO Discoveries from the {\it New Horizons} Deep Drilling Field.\label{tab:timetable}}

        \tablehead{
        \colhead{Search Depth} &
        \colhead{$< $1~au} &
        \colhead{Obs. by NH} &
        \colhead{Total} &
        \colhead{Rubin Time} \\
        \colhead{r (5-$\sigma$)} &
        \colhead{No.} &
        \colhead{No.} &
        \colhead{No.} &
        \colhead{Hours\tablenotemark{a}}
}

\decimals

\startdata 
27.0 & 1.2 & 6 &  450  & 12\\ 
27.5 & 2.9 & 12 & 780  & 30\\ 
28.0 & 4.6 & 22 & 1280 & 75\\ 
28.5 & 7.6 & 38 & 2000 & 180\\
\enddata
\tablenotetext{a}{Total time required to enable discovery and first-order orbit determination, six visits over one year.}
\end{deluxetable*}

\subsection{Analysis Techniques}

\begin{figure}[ht]
\plotone{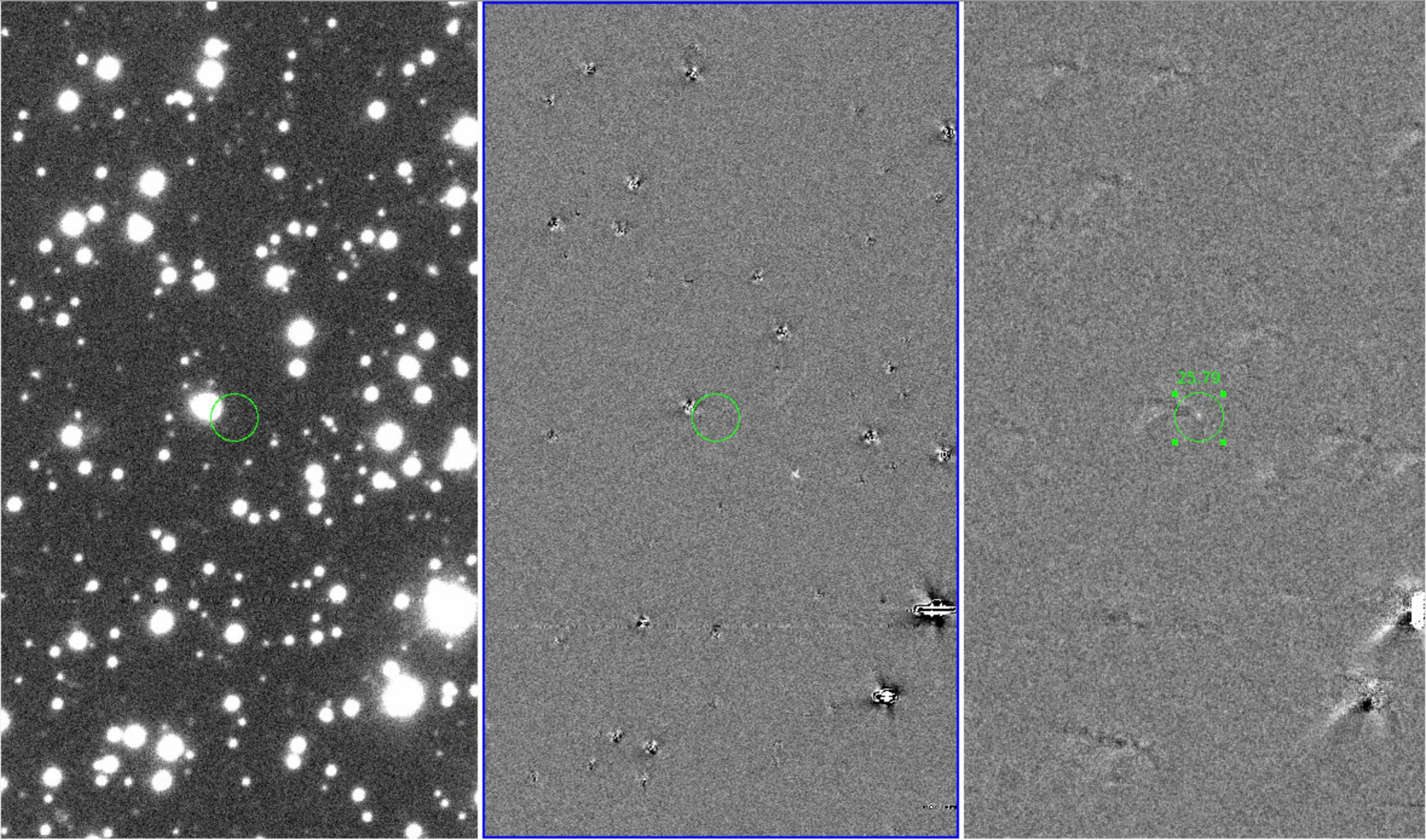}
\caption{\label{fig-sample}
Sample of image series stacks followed by subtraction of a sky template of stable sources and then identification of moving objects within those differenced images. 
}
\end{figure}

\begin{figure}[ht]
\plotone{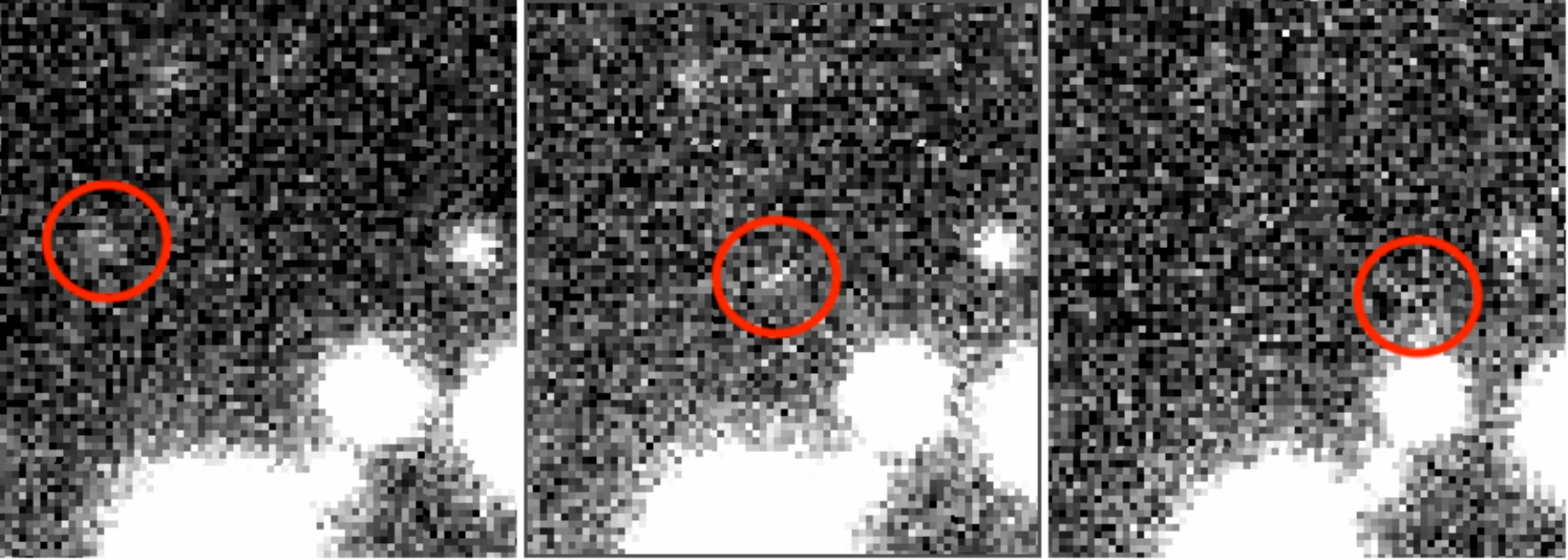}
\plotone{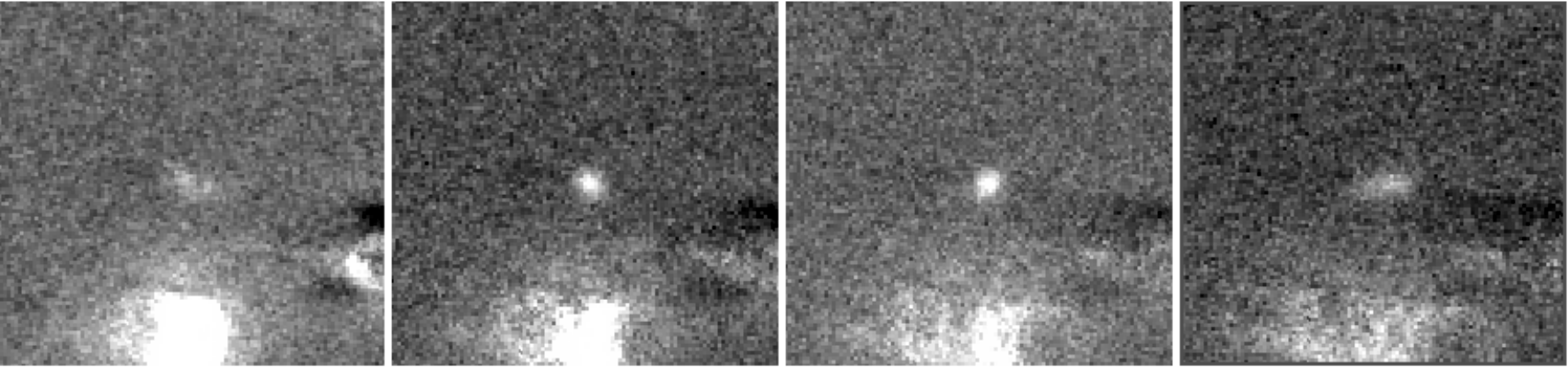}
\caption{\label{fig-shiftnstack} Example of detection of a real moving object made via the shift-and-stack technique from \citet{Fraser2024PSJ}. \textbf{Top:} three individual frames acquired with the Hyper Suprime-Cam on the Subaru Telescope. Red circles mark the object's positions, barely visible above the noise floor. \textbf{Bottom:} shift-stacks of difference imagery from the same regions as presented above. The stacks were made at rates of motion of 0.5, 1.0, 1.5, and 2.0 $\arcsec$/hr (left to right). The true rate of motion of the object is 1.9 $\arcsec$/hr. }
\end{figure}

Because Pluto was traversing the galactic plane when {\it New Horizons} flew past it, most of the search region for spacecraft flyby targets have been in highly populated stellar fields. This Rubin survey must, therefore, also be carried out near the galactic plane ($b \sim -14.8^\circ$) and at a position along the {\it New Horizons} trajectory because it allows both characterization of the deep Kuiper Belt and the potential discovery of one final flyby object for the spacecraft (Figure \ref{fig-NH_Placement}). 
Techniques to stack sets of short exposures (see Figure~\ref{fig-shiftnstack}) collected close to each other in time and then compared with later sets of close-in-time exposures to track these objects and determine their orbits have been developed \citep{Fraser2024LPI, Yoshida2024}.
\textbf{The technique does not require a template as the reference image used for differencing is built from the micro-survey observations.} The field shifts between nights are blind to the detection process (i.e., the detections from one night are not used to plan future fields), and the template is built from the micro-survey observations combined across multiple visits. 
Over time, highly effective machine learning techniques have also been developed to cope with these dense sky backgrounds and false positives (Figure \ref{fig-sample}; see also \citet{Napier2023,Buie2024,Fraser2024PSJ}). 
These techniques will be applied to search for KBOs in the the proposed micro-survey, but they can also be used to search for other small body populations in the NHDDF.

\section{Summary}
This white paper proposes an extremely deep Rubin micro-survey in the {\it New Horizons} trajectory field (the New Horizons Deep Drilling Field) to both look for spacecraft-accessible KBOs for close flyby or distant study at high solar phase angles and possibly at high spatial resolution and to investigate fundamental properties of the distant solar system. As designed, this survey, using 30 hours of Rubin time spread over six 5-hour visits (4 spanning two lunations in a first year, two in the following year), will discover and determine orbits for as many as 730 KBOs to a Rubin $r$ magnitude of 27.5, providing a unique opportunity for ground-breaking Kuiper Belt science specifically related to the size distribution. It may discover as many as 12 KBOs observable with the {\it New Horizons} LORRI camera and $\sim$3 objects within 1~au of the spacecraft. Rapid follow-up of particularly interesting Rubin discoveries with HST or JWST could enable {\it New Horizons} observations as early as one year after the first opposition visits. 

\section{Acknowledgments}
We thank NASA for funding and continued support of the New Horizons mission; this work was supported by the {\it New Horizons} Project (NASW-02008).
This research has made use of NASA’s Astrophysics Data System Bibliographic Services. This research has used data and/or services provided by the International Astronomical Union's Minor Planet Center. Some of the results in this paper have been derived using the healpy and HEALPix\footnote{http://healpix.sf.net} packages.

\software{Sorcha (Merritt et al. {et al., \it submitted}, Holman {et al. \it submitted}), ASSIST \citep{2023PSJ.....4...69H,hanno_rein_2023_7778017},Astropy \citep{2013A&A...558A..33A,2018AJ....156..123A,2022ApJ...935..167A}, Healpy \citep{Zonca2019,2005ApJ...622..759G}, Matplotlib \citep{Hunter:2007}, Numba \citep{2015llvm.confE...1L}, Numpy \citep{harris2020array}, pandas \citep{mckinney-proc-scipy-2010, reback2020pandas}, Pooch \citep{uieda2020}, PyTables \citep{pytables}, REBOUND \citep{rebound,reboundias15}, rubin$\_$sim \citep{2018Icar..303..181J,peter_yoachim_2022_7087823}, sbpy \citep{2019JOSS....4.1426M}, SciPy \citep{2020SciPy-NMeth}, Spiceypy \citep{2020JOSS....5.2050A}, sqlite (\url{https://www.sqlite.org/index.html}), sqlite3 (\url{https://docs.python.org/3/library/sqlite3.html}), tqdm \citep{casper_da_costa_luis_2023_8233425}, Black (\url{https://black.readthedocs.io/en/stable/faq.html}), Jupyter Notebooks \citep{soton403913}}

\bibliographystyle{aasjournal}
\bibliography{NHKBOs_v1}

\begin{thebibliography}{}
\expandafter\ifx\csname natexlab\endcsname\relax\def\natexlab#1{#1}\fi
\providecommand{\url}[1]{\href{#1}{#1}}
\providecommand{\dodoi}[1]{doi:~\href{http://doi.org/#1}{\nolinkurl{#1}}}
\providecommand{\doeprint}[1]{\href{http://ascl.net/#1}{\nolinkurl{http://ascl.net/#1}}}
\providecommand{\doarXiv}[1]{\href{https://arxiv.org/abs/#1}{\nolinkurl{https://arxiv.org/abs/#1}}}

\bibitem[{{Annex} {et~al.}(2020){Annex}, {Pearson}, {Seignovert}, {Carcich},
  {Eichhorn}, {Mapel}, {von Forstner}, {McAuliffe}, {del Rio}, {Berry}, {Aye},
  {Stefko}, {de Val-Borro}, {Kulumani}, \& {Murakami}}]{2020JOSS....5.2050A}
{Annex}, A., {Pearson}, B., {Seignovert}, B., {et~al.} 2020, The Journal of
  Open Source Software, 5, 2050, \dodoi{10.21105/joss.02050}

\bibitem[{{Astropy Collaboration} {et~al.}(2013){Astropy Collaboration},
  {Robitaille}, {Tollerud}, {Greenfield}, {Droettboom}, {Bray}, {Aldcroft},
  {Davis}, {Ginsburg}, {Price-Whelan}, {Kerzendorf}, {Conley}, {Crighton},
  {Barbary}, {Muna}, {Ferguson}, {Grollier}, {Parikh}, {Nair}, {Unther},
  {Deil}, {Woillez}, {Conseil}, {Kramer}, {Turner}, {Singer}, {Fox}, {Weaver},
  {Zabalza}, {Edwards}, {Azalee Bostroem}, {Burke}, {Casey}, {Crawford},
  {Dencheva}, {Ely}, {Jenness}, {Labrie}, {Lim}, {Pierfederici}, {Pontzen},
  {Ptak}, {Refsdal}, {Servillat}, \& {Streicher}}]{2013A&A...558A..33A}
{Astropy Collaboration}, {Robitaille}, T.~P., {Tollerud}, E.~J., {et~al.} 2013,
  \aap, 558, A33, \dodoi{10.1051/0004-6361/201322068}

\bibitem[{{Astropy Collaboration} {et~al.}(2018){Astropy Collaboration},
  {Price-Whelan}, {Sip{\H{o}}cz}, {G{\"u}nther}, {Lim}, {Crawford}, {Conseil},
  {Shupe}, {Craig}, {Dencheva}, {Ginsburg}, {VanderPlas}, {Bradley},
  {P{'e}rez-Su{'a}rez}, {de Val-Borro}, {Aldcroft}, {Cruz}, {Robitaille},
  {Tollerud}, {Ardelean}, {Babej}, {Bach}, {Bachetti}, {Bakanov}, {Bamford},
  {Barentsen}, {Barmby}, {Baumbach}, {Berry}, {Biscani}, {Boquien}, {Bostroem},
  {Bouma}, {Brammer}, {Bray}, {Breytenbach}, {Buddelmeijer}, {Burke},
  {Calderone}, {Cano Rodr{\'\i}guez}, {Cara}, {Cardoso}, {Cheedella}, {Copin},
  {Corrales}, {Crichton}, {D'Avella}, {Deil}, {Depagne}, {Dietrich}, {Donath},
  {Droettboom}, {Earl}, {Erben}, {Fabbro}, {Ferreira}, {Finethy}, {Fox},
  {Garrison}, {Gibbons}, {Goldstein}, {Gommers}, {Greco}, {Greenfield},
  {Groener}, {Grollier}, {Hagen}, {Hirst}, {Homeier}, {Horton}, {Hosseinzadeh},
  {Hu}, {Hunkeler}, {Ivezi{\'c}}, {Jain}, {Jenness}, {Kanarek}, {Kendrew},
  {Kern}, {Kerzendorf}, {Khvalko}, {King}, {Kirkby}, {Kulkarni}, {Kumar},
  {Lee}, {Lenz}, {Littlefair}, {Ma}, {Macleod}, {Mastropietro}, {McCully},
  {Montagnac}, {Morris}, {Mueller}, {Mumford}, {Muna}, {Murphy}, {Nelson},
  {Nguyen}, {Ninan}, {N{"o}the}, {Ogaz}, {Oh}, {Parejko}, {Parley}, {Pascual},
  {Patil}, {Patil}, {Plunkett}, {Prochaska}, {Rastogi}, {Reddy Janga},
  {Sabater}, {Sakurikar}, {Seifert}, {Sherbert}, {Sherwood-Taylor}, {Shih},
  {Sick}, {Silbiger}, {Singanamalla}, {Singer}, {Sladen}, {Sooley},
  {Sornarajah}, {Streicher}, {Teuben}, {Thomas}, {Tremblay}, {Turner},
  {Terr{\'o}n}, {van Kerkwijk}, {de la Vega}, {Watkins}, {Weaver}, {Whitmore},
  {Woillez}, {Zabalza}, \& {Astropy Contributors}}]{2018AJ....156..123A}
{Astropy Collaboration}, {Price-Whelan}, A.~M., {Sip{\H{o}}cz}, B.~M., {et~al.}
  2018, \aj, 156, 123, \dodoi{10.3847/1538-3881/aabc4f}

\bibitem[{{Astropy Collaboration} {et~al.}(2022){Astropy Collaboration},
  {Price-Whelan}, {Lim}, {Earl}, {Starkman}, {Bradley}, {Shupe}, {Patil},
  {Corrales}, {Brasseur}, {N{\"o}the}, {Donath}, {Tollerud}, {Morris},
  {Ginsburg}, {Vaher}, {Weaver}, {Tocknell}, {Jamieson}, {van Kerkwijk},
  {Robitaille}, {Merry}, {Bachetti}, {G{\"u}nther}, {Aldcroft},
  {Alvarado-Montes}, {Archibald}, {B{'o}di}, {Bapat}, {Barentsen}, {Baz{'a}n},
  {Biswas}, {Boquien}, {Burke}, {Cara}, {Cara}, {Conroy}, {Conseil}, {Craig},
  {Cross}, {Cruz}, {D'Eugenio}, {Dencheva}, {Devillepoix}, {Dietrich},
  {Eigenbrot}, {Erben}, {Ferreira}, {Foreman-Mackey}, {Fox}, {Freij}, {Garg},
  {Geda}, {Glattly}, {Gondhalekar}, {Gordon}, {Grant}, {Greenfield}, {Groener},
  {Guest}, {Gurovich}, {Handberg}, {Hart}, {Hatfield-Dodds}, {Homeier},
  {Hosseinzadeh}, {Jenness}, {Jones}, {Joseph}, {Kalmbach}, {Karamehmetoglu},
  {Ka{\l}uszy{'n}ski}, {Kelley}, {Kern}, {Kerzendorf}, {Koch}, {Kulumani},
  {Lee}, {Ly}, {Ma}, {MacBride}, {Maljaars}, {Muna}, {Murphy}, {Norman},
  {O'Steen}, {Oman}, {Pacifici}, {Pascual}, {Pascual-Granado}, {Patil},
  {Perren}, {Pickering}, {Rastogi}, {Roulston}, {Ryan}, {Rykoff}, {Sabater},
  {Sakurikar}, {Salgado}, {Sanghi}, {Saunders}, {Savchenko}, {Schwardt},
  {Seifert-Eckert}, {Shih}, {Jain}, {Shukla}, {Sick}, {Simpson},
  {Singanamalla}, {Singer}, {Singhal}, {Sinha}, {Sip{\H{o}}cz}, {Spitler},
  {Stansby}, {Streicher}, {{\v{S}}umak}, {Swinbank}, {Taranu}, {Tewary},
  {Tremblay}, {de Val-Borro}, {Van Kooten}, {Vasovi{\'c}}, {Verma}, {de Miranda
  Cardoso}, {Williams}, {Wilson}, {Winkel}, {Wood-Vasey}, {Xue}, {Yoachim},
  {Zhang}, {Zonca}, \& {Astropy Project Contributors}}]{2022ApJ...935..167A}
{Astropy Collaboration}, {Price-Whelan}, A.~M., {Lim}, P.~L., {et~al.} 2022,
  \apj, 935, 167, \dodoi{10.3847/1538-4357/ac7c74}

\bibitem[{{Bannister} {et~al.}(2016){Bannister}, {Kavelaars}, {Petit},
  {Gladman}, {Gwyn}, {Chen}, {Volk}, {Alexandersen}, {Benecchi}, {Delsanti},
  {Fraser}, {Granvik}, {Grundy}, {Guilbert-Lepoutre}, {Hestroffer}, {Ip},
  {Jakubik}, {Jones}, {Kaib}, {Kavelaars}, {Lacerda}, {Lawler}, {Lehner},
  {Lin}, {Lister}, {Lykawka}, {Monty}, {Marsset}, {Murray-Clay}, {Noll},
  {Parker}, {Pike}, {Rousselot}, {Rusk}, {Schwamb}, {Shankman}, {Sicardy},
  {Vernazza}, \& {Wang}}]{Bannister2016}
{Bannister}, M.~T., {Kavelaars}, J.~J., {Petit}, J.-M., {et~al.} 2016, \aj,
  152, 70, \dodoi{10.3847/0004-6256/152/3/70}

\bibitem[{{Bannister} {et~al.}(2018){Bannister}, {Gladman}, {Kavelaars},
  {Petit}, {Volk}, {Chen}, {Alexandersen}, {Gwyn}, {Schwamb}, {Ashton},
  {Benecchi}, {Cabral}, {Dawson}, {Delsanti}, {Fraser}, {Granvik},
  {Greenstreet}, {Guilbert-Lepoutre}, {Ip}, {Jakubik}, {Jones}, {Kaib},
  {Lacerda}, {Van Laerhoven}, {Lawler}, {Lehner}, {Lin}, {Lykawka}, {Marsset},
  {Murray-Clay}, {Pike}, {Rousselot}, {Shankman}, {Thirouin}, {Vernazza}, \&
  {Wang}}]{Bannister2018}
{Bannister}, M.~T., {Gladman}, B.~J., {Kavelaars}, J.~J., {et~al.} 2018, \apjs,
  236, 18, \dodoi{10.3847/1538-4365/aab77a}

\bibitem[{{Beaudoin} {et~al.}(2023){Beaudoin}, {Gladman}, {Huang}, {Bannister},
  {Kavelaars}, {Petit}, \& {Volk}}]{beaudoin2023}
{Beaudoin}, M., {Gladman}, B., {Huang}, Y., {et~al.} 2023, \psj, 4, 145,
  \dodoi{10.3847/PSJ/ace88d}

\bibitem[{{Benecchi} \& {Sheppard}(2013)}]{Benechhi13}
{Benecchi}, S.~D., \& {Sheppard}, S.~S. 2013, \aj, 145, 124,
  \dodoi{10.1088/0004-6256/145/5/124}

\bibitem[{{Bernstein} {et~al.}(2004){Bernstein}, {Trilling}, {Allen}, {Brown},
  {Holman}, \& {Malhotra}}]{Bernstein2004}
{Bernstein}, G.~M., {Trilling}, D.~E., {Allen}, R.~L., {et~al.} 2004, \aj, 128,
  1364, \dodoi{10.1086/422919}

\bibitem[{{Bianco F. and The Rubin Observatory Survey Cadence Optimization
  Committee}(2023)}]{PSTN-055}
{Bianco F. and The Rubin Observatory Survey Cadence Optimization Committee}.
  2023, {Survey Cadence Optimization Committee’s Phase 2 Recommendations},
  {Vera C. Rubin Observatory}.
\newblock \url{https://pstn-055.lsst.io/}

\bibitem[{{Bianco F. and The Rubin Observatory Survey Cadence Optimization
  Committee}(2024)}]{PSTN-056}
---. 2024, {Survey Cadence Optimization Committee’s Phase 3 Recommendations},
   {Vera C. Rubin Observatory}.
\newblock \url{https://pstn-056.lsst.io/}

\bibitem[{{Bianco, Federica B. and Jones, Lynne and Ivezi\'{c}, {\v Z}eljko and
  Ritz, Steven and the Rubin Project Science Team}(2022)}]{PSTN-054}
{Bianco, Federica B. and Jones, Lynne and Ivezi\'{c}, {\v Z}eljko and Ritz,
  Steven and the Rubin Project Science Team}. 2022, {Updated estimates of the
  Rubin system throughput and expected LSST image depth},  {Vera C. Rubin
  Observatory}.
\newblock \url{https://pstn-054.lsst.io/}

\bibitem[{{Bonnarel} {et~al.}(2000){Bonnarel}, {Fernique}, {Bienaym{\'e}},
  {Egret}, {Genova}, {Louys}, {Ochsenbein}, {Wenger}, \& {Bartlett}}]{ALADIN}
{Bonnarel}, F., {Fernique}, P., {Bienaym{\'e}}, O., {et~al.} 2000, \aaps, 143,
  33, \dodoi{10.1051/aas:2000331}

\bibitem[{{Buie} \& {Keller}(2016)}]{RECON16}
{Buie}, M.~W., \& {Keller}, J.~M. 2016, \aj, 151, 73,
  \dodoi{10.3847/0004-6256/151/3/73}

\bibitem[{{Buie} {et~al.}(2024){Buie}, {Spencer}, {Porter}, {Benecchi},
  {Parker}, {Stern}, {Belton}, {Binzel}, {Borncamp}, {DeMeo}, {Fabbro},
  {Fuentes}, {Furusawa}, {Fuse}, {Gay}, {Gwyn}, {Holman}, {Karoji},
  {Kavelaars}, {Kinoshita}, {Miyazaki}, {Mountain}, {Noll}, {Osip}, {Petit},
  {Reid}, {Sheppard}, {Showalter}, {Steffl}, {Sterner}, {Tajitsu}, {Tholen},
  {Trilling}, {Weaver}, {Verbiscer}, {Wasserman}, {Yamashita}, {Yanagisawa},
  {Yoshida}, \& {Zangari}}]{Buie2024}
{Buie}, M.~W., {Spencer}, J.~R., {Porter}, S.~B., {et~al.} 2024, \psj, 5, 196,
  \dodoi{10.3847/PSJ/ad676d}

\bibitem[{{Cheng} {et~al.}(2008){Cheng}, {Weaver}, {Conard}, {Morgan},
  {Barnouin-Jha}, {Boldt}, {Cooper}, {Darlington}, {Grey}, {Hayes},
  {Kosakowski}, {Magee}, {Rossano}, {Sampath}, {Schlemm}, \&
  {Taylor}}]{2008SSRv..140..189C}
{Cheng}, A.~F., {Weaver}, H.~A., {Conard}, S.~J., {et~al.} 2008, \ssr, 140,
  189, \dodoi{10.1007/s11214-007-9271-6}

\bibitem[{{Chiang} \& {Brown}(1999)}]{Chiang1999}
{Chiang}, E.~I., \& {Brown}, M.~E. 1999, \aj, 118, 1411, \dodoi{10.1086/301005}

\bibitem[{{Crompvoets} {et~al.}(2022){Crompvoets}, {Lawler}, {Volk}, {Chen},
  {Gladman}, {Peltier}, {Alexandersen}, {Bannister}, {Gwyn}, {Kavelaars}, \&
  {Petit}}]{Crompvoets2022}
{Crompvoets}, B.~L., {Lawler}, S.~M., {Volk}, K., {et~al.} 2022, \psj, 3, 113,
  \dodoi{10.3847/PSJ/ac67e0}

\bibitem[{da~Costa-Luis {et~al.}(2023)da~Costa-Luis, Larroque, Altendorf, Mary,
  richardsheridan, Korobov, Yorav-Raphael, Ivanov, Bargull, Rodrigues, Chen,
  Lee, Newey, CrazyPython, JC, Zugnoni, Pagel, mjstevens777, Dektyarev,
  Rothberg, Plavin, Dill, FichteFoll, Sturm, HeoHeo, van Kemenade, McCracken,
  MapleCCC, Nordlund, \& Boyle}]{casper_da_costa_luis_2023_8233425}
da~Costa-Luis, C., Larroque, S.~K., Altendorf, K., {et~al.} 2023, {tqdm: A
  fast, Extensible Progress Bar for Python and CLI}, v4.66.1,  Zenodo,
  \dodoi{10.5281/zenodo.8233425}

\bibitem[{Desch \& Neveu(2017)}]{DESCH2017}
Desch, S., \& Neveu, M. 2017, Icarus, 287, 175,
  \dodoi{https://doi.org/10.1016/j.icarus.2016.11.037}

\bibitem[{{Doner} {et~al.}(2024){Doner}, {Hor{\'a}nyi}, {Bagenal}, {Brandt},
  {Grundy}, {Lisse}, {Parker}, {Poppe}, {Singer}, {Stern}, \&
  {Verbiscer}}]{Doner2024}
{Doner}, A., {Hor{\'a}nyi}, M., {Bagenal}, F., {et~al.} 2024, \apjl, 961, L38,
  \dodoi{10.3847/2041-8213/ad18b0}

\bibitem[{{Fraser} {et~al.}(2024{\natexlab{a}}){Fraser}, {Porter}, {Peltier},
  {Kavelaars}, {Verbiscer}, {Buie}, {Stern}, {Spencer}, {Benecchi}, {Terai},
  {Ito}, {Yoshida}, {Gerdes}, {Napier}, {Lin}, {Gwyn}, {Smotherman}, {Fabbro},
  {Singer}, {Alexander}, {Arimatsu}, {Banks}, {Bray}, {Ramy El-Maarry},
  {Ferrell}, {Fuse}, {Glass}, {Holt}, {Hong}, {Ishimaru}, {Johnson}, {Lauer},
  {Leiva}, {S. Lykawka}, {Marschall}, {N{\'u}{\~n}ez}, {Postman}, {Quirico},
  {Rhoden}, {Simpson}, {Schenk}, {Skrutskie}, {Steffl}, \&
  {Throop}}]{Fraser2024PSJ}
{Fraser}, W.~C., {Porter}, S.~B., {Peltier}, L., {et~al.} 2024{\natexlab{a}},
  \psj, 5, 227, \dodoi{10.3847/PSJ/ad6f9e}

\bibitem[{{Fraser} {et~al.}(2024{\natexlab{b}}){Fraser}, {Porter}, {Benecchi},
  {Kavelaars}, {Verbiscer}, {Weaver}, {Spencer}, {Buie}, {Yoshida}, {Ito},
  {Terai}, {Stern}, {Parker}, {Singer}, {Brandt}, {Peltier}, {Gerdes}, {Lin},
  {Naiper}, {New Horizons KBO Search}, \& {Planetary Teams}}]{Fraser2024LPI}
{Fraser}, W.~C., {Porter}, S.~B., {Benecchi}, S.~D., {et~al.}
  2024{\natexlab{b}}, in LPI Contributions, Vol. 3040, LPI Contributions, 2440

\bibitem[{{Gladman} \& {Volk}(2021)}]{Gladman2022}
{Gladman}, B., \& {Volk}, K. 2021, \araa, 59, 203,
  \dodoi{10.1146/annurev-astro-120920-010005}

\bibitem[{{G{'o}rski} {et~al.}(2005){G{'o}rski}, {Hivon}, {Banday}, {Wandelt},
  {Hansen}, {Reinecke}, \& {Bartelmann}}]{2005ApJ...622..759G}
{G{'o}rski}, K.~M., {Hivon}, E., {Banday}, A.~J., {et~al.} 2005, \apj, 622,
  759, \dodoi{10.1086/427976}

\bibitem[{{Grundy} {et~al.}(2020){Grundy}, {Bird}, {Britt}, {Cook},
  {Cruikshank}, {Howett}, {Krijt}, {Linscott}, {Olkin}, {Parker}, {Protopapa},
  {Ruaud}, {Umurhan}, {Young}, {Dalle Ore}, {Kavelaars}, {Keane}, {Pendleton},
  {Porter}, {Scipioni}, {Spencer}, {Stern}, {Verbiscer}, {Weaver}, {Binzel},
  {Buie}, {Buratti}, {Cheng}, {Earle}, {Elliott}, {Gabasova}, {Gladstone},
  {Hill}, {Horanyi}, {Jennings}, {Lunsford}, {McComas}, {McKinnon}, {McNutt},
  {Moore}, {Parker}, {Quirico}, {Reuter}, {Schenk}, {Schmitt}, {Showalter},
  {Singer}, {Weigle}, \& {Zangari}}]{Grundy2020}
{Grundy}, W.~M., {Bird}, M.~K., {Britt}, D.~T., {et~al.} 2020, Science, 367,
  aay3705, \dodoi{10.1126/science.aay3705}

\bibitem[{Harris {et~al.}(2020)Harris, Millman, van~der Walt, Gommers,
  Virtanen, Cournapeau, Wieser, Taylor, Berg, Smith, Kern, Picus, Hoyer, van
  Kerkwijk, Brett, Haldane, del R{\'{i}}o, Wiebe, Peterson,
  G{\'{e}}rard-Marchant, Sheppard, Reddy, Weckesser, Abbasi, Gohlke, \&
  Oliphant}]{harris2020array}
Harris, C.~R., Millman, K.~J., van~der Walt, S.~J., {et~al.} 2020, Nature, 585,
  357, \dodoi{10.1038/s41586-020-2649-2}

\bibitem[{{Holman} {et~al.}(2023){Holman}, {Akmal}, {Farnocchia}, {Rein},
  {Payne}, {Weryk}, {Tamayo}, \& {Hernandez}}]{2023PSJ.....4...69H}
{Holman}, M.~J., {Akmal}, A., {Farnocchia}, D., {et~al.} 2023, \psj, 4, 69,
  \dodoi{10.3847/PSJ/acc9a9}

\bibitem[{Hunter(2007)}]{Hunter:2007}
Hunter, J.~D. 2007, Computing in Science \& Engineering, 9, 90,
  \dodoi{10.1109/MCSE.2007.55}

\bibitem[{{Jones} {et~al.}(2018){Jones}, {Slater}, {Moeyens}, {Allen},
  {Axelrod}, {Cook}, {Ivezi{\'c}}, {Juri{\'c}}, {Myers}, \&
  {Petry}}]{2018Icar..303..181J}
{Jones}, R.~L., {Slater}, C.~T., {Moeyens}, J., {et~al.} 2018, \icarus, 303,
  181, \dodoi{10.1016/j.icarus.2017.11.033}

\bibitem[{Kavelaars {et~al.}(2020)Kavelaars, Lawler, Bannister, \&
  Shankman}]{Kavelaars2020}
Kavelaars, J., Lawler, S.~M., Bannister, M.~T., \& Shankman, C. 2020, in The
  Trans-Neptunian Solar System, ed. D.~Prialnik, M.~A. Barucci, \& L.~A. Young
  (Elsevier), 61--77,
  \dodoi{https://doi.org/10.1016/B978-0-12-816490-7.00003-5}

\bibitem[{{Kavelaars} {et~al.}(2021){Kavelaars}, {Petit}, {Gladman},
  {Bannister}, {Alexandersen}, {Chen}, {Gwyn}, \& {Volk}}]{Kavelaars2021}
{Kavelaars}, J.~J., {Petit}, J.-M., {Gladman}, B., {et~al.} 2021, \apjl, 920,
  L28, \dodoi{10.3847/2041-8213/ac2c72}

\bibitem[{Kluyver {et~al.}(2016)Kluyver, Ragan-Kelley, P{\'e}rez, Granger,
  Bussonnier, Frederic, Kelley, Hamrick, Grout, Corlay, Ivanov, Avila, Abdalla,
  Willing, \& development team}]{soton403913}
Kluyver, T., Ragan-Kelley, B., P{\'e}rez, F., {et~al.} 2016, in Positioning and
  Power in Academic Publishing: Players, Agents and Agendas (IOS Press),
  87--90.
\newblock \url{https://eprints.soton.ac.uk/403913/}

\bibitem[{Krasnopolsky(2020)}]{KRASNOPOLSKY2020}
Krasnopolsky, V.~A. 2020, Icarus, 335, 113374,
  \dodoi{https://doi.org/10.1016/j.icarus.2019.07.008}

\bibitem[{{Lam} {et~al.}(2015){Lam}, {Pitrou}, \&
  {Seibert}}]{2015llvm.confE...1L}
{Lam}, S.~K., {Pitrou}, A., \& {Seibert}, S. 2015, in Proc. Second Workshop on
  the LLVM Compiler Infrastructure in HPC, 1--6,
  \dodoi{10.1145/2833157.2833162}

\bibitem[{{Lawler} {et~al.}(2018){Lawler}, {Kavelaars}, {Alexandersen},
  {Bannister}, {Gladman}, {Petit}, \& {Shankman}}]{Lawler2018}
{Lawler}, S.~M., {Kavelaars}, J.~J., {Alexandersen}, M., {et~al.} 2018,
  Frontiers in Astronomy and Space Sciences, 5, 14,
  \dodoi{10.3389/fspas.2018.00014}

\bibitem[{{Lawler} \& {Pike}(2024)}]{Lawler24}
{Lawler}, S.~M., \& {Pike}, R.~E. 2024, arXiv e-prints, arXiv:2410.04338,
  \dodoi{10.48550/arXiv.2410.04338}

\bibitem[{{McKinnon} {et~al.}(2020){McKinnon}, {Richardson}, {Marohnic},
  {Keane}, {Grundy}, {Hamilton}, {Nesvorn{\'y}}, {Umurhan}, {Lauer}, {Singer},
  {Stern}, {Weaver}, {Spencer}, {Buie}, {Moore}, {Kavelaars}, {Lisse}, {Mao},
  {Parker}, {Porter}, {Showalter}, {Olkin}, {Cruikshank}, {Elliott},
  {Gladstone}, {Parker}, {Verbiscer}, {Young}, \& {New Horizons Science
  Team}}]{McKinnon2020}
{McKinnon}, W.~B., {Richardson}, D.~C., {Marohnic}, J.~C., {et~al.} 2020,
  Science, 367, aay6620, \dodoi{10.1126/science.aay6620}

\bibitem[{{Mommert} {et~al.}(2019){Mommert}, {Kelley}, {de Val-Borro}, {Li},
  {Guzman}, {Sip{\H{o}}cz}, {{\v{D}}urech}, {Granvik}, {Grundy}, {Moskovitz},
  {Penttil{\"a}}, \& {Samarasinha}}]{2019JOSS....4.1426M}
{Mommert}, M., {Kelley}, M., {de Val-Borro}, M., {et~al.} 2019, The Journal of
  Open Source Software, 4, 1426, \dodoi{10.21105/joss.01426}

\bibitem[{{Napier} {et~al.}(2023){Napier}, {Lin}, {Gerdes}, {Adams}, {Simpson},
  {Porter}, {Weber}, {Markwardt}, {Gowman}, {Smotherman}, {Bernardinelli},
  {Juri{\'c}}, {Connolly}, {Bryce Kalmbach}, {Portillo}, {Trilling}, {Strauss},
  {Oldroyd}, {Trujillo}, {Chandler}, {Holman}, {Schlichting}, {McNeill}, \&
  {the DEEP Collaboration}}]{Napier2023}
{Napier}, K.~J., {Lin}, H.-W., {Gerdes}, D.~W., {et~al.} 2023, arXiv e-prints,
  arXiv:2309.09478, \dodoi{10.48550/arXiv.2309.09478}

\bibitem[{{Nesvorn{\'y}} {et~al.}(2023{\natexlab{a}}){Nesvorn{\'y}},
  {Bernardinelli}, {Vokrouhlick{\'y}}, \& {Batygin}}]{Nesvorny2023}
{Nesvorn{\'y}}, D., {Bernardinelli}, P., {Vokrouhlick{\'y}}, D., \& {Batygin},
  K. 2023{\natexlab{a}}, \icarus, 406, 115738,
  \dodoi{10.1016/j.icarus.2023.115738}

\bibitem[{{Nesvorn{\'y}} {et~al.}(2023{\natexlab{b}}){Nesvorn{\'y}}, {Dones},
  {De Pr{\'a}}, {Womack}, \& {Zahnle}}]{Nesvorny2023PSJ}
{Nesvorn{\'y}}, D., {Dones}, L., {De Pr{\'a}}, M., {Womack}, M., \& {Zahnle},
  K.~J. 2023{\natexlab{b}}, \psj, 4, 139, \dodoi{10.3847/PSJ/ace8ff}

\bibitem[{{Nesvorn{\'y}} {et~al.}(2021){Nesvorn{\'y}}, {Li}, {Simon}, {Youdin},
  {Richardson}, {Marschall}, \& {Grundy}}]{Nesvorny2021}
{Nesvorn{\'y}}, D., {Li}, R., {Simon}, J.~B., {et~al.} 2021, \psj, 2, 27,
  \dodoi{10.3847/PSJ/abd858}

\bibitem[{{Nesvorn{\'y}} {et~al.}(2022){Nesvorn{\'y}}, {Vokrouhlick{\'y}}, \&
  {Fraser}}]{Nesvorny2022}
{Nesvorn{\'y}}, D., {Vokrouhlick{\'y}}, D., \& {Fraser}, W.~C. 2022, \aj, 163,
  137, \dodoi{10.3847/1538-3881/ac4bc9}

\bibitem[{pandas~development team(2020)}]{reback2020pandas}
pandas~development team, T. 2020, pandas-dev/pandas: Pandas, latest,  Zenodo,
  \dodoi{10.5281/zenodo.3509134}

\bibitem[{{Peltier} {et~al.}(2022){Peltier}, {Kavelaars}, {Fraser}, {Porter},
  {Lawler}, \& {The New Horizons Team}}]{Peltier2022}
{Peltier}, L., {Kavelaars}, J., {Fraser}, W., {et~al.} 2022, in AAS/Division
  for Planetary Sciences Meeting Abstracts, Vol.~54, AAS/Division for Planetary
  Sciences Meeting Abstracts, 501.09

\bibitem[{{Petit} {et~al.}(2023{\natexlab{a}}){Petit}, {Gladman}, {Kavelaars},
  {Bannister}, {Alexandersen}, {Volk}, \& {Chen}}]{Petit2023}
{Petit}, J.-M., {Gladman}, B., {Kavelaars}, J.~J., {et~al.} 2023{\natexlab{a}},
  \apjl, 947, L4, \dodoi{10.3847/2041-8213/acc525}

\bibitem[{{Petit} {et~al.}(2023{\natexlab{b}}){Petit}, {Gladman}, {Kavelaars},
  {Volk}, {Crompvoets}, {Lawler}, {Beaudoin}, {Peltier}, {Bannister},
  {Alexandersen}, {Chen}, {Gwin}, \& {Kaib}}]{Petit2023DPS}
{Petit}, J.-M., {Gladman}, B., {Kavelaars}, J., {et~al.} 2023{\natexlab{b}}, in
  AAS/Division for Planetary Sciences Meeting Abstracts, Vol.~55, AAS/Division
  for Planetary Sciences Meeting Abstracts \#55, 209.04

\bibitem[{{Polak} \& {Klahr}(2023)}]{Polak2023}
{Polak}, B., \& {Klahr}, H. 2023, \apj, 943, 125,
  \dodoi{10.3847/1538-4357/aca58f}

\bibitem[{{Porter} {et~al.}(2024){Porter}, {Singer}, {Schenk}, {Verbiscer},
  {Grundy}, {Benecchi}, {Parker}, {Brandt}, \& {Stern}}]{Porter2024}
{Porter}, S., {Singer}, K.~N., {Schenk}, P., {et~al.} 2024, in AGU Meeting
  Abstracts, Vol. 2024, P24B--03.
\newblock \url{https://agu.confex.com/agu/agu24/meetingapp.cgi/Paper/1679982}

\bibitem[{{Porter} {et~al.}(2016){Porter}, {Spencer}, {Benecchi}, {Verbiscer},
  {Zangari}, {Weaver}, {Lauer}, {Parker}, {Buie}, {Cheng}, {Young}, {Olkin},
  {Ennico}, {Stern}, \& {New Horizons Science Team}}]{Porter2016}
{Porter}, S.~B., {Spencer}, J.~R., {Benecchi}, S., {et~al.} 2016, \apjl, 828,
  L15, \dodoi{10.3847/2041-8205/828/2/L15}

\bibitem[{Rein {et~al.}(2023)Rein, Holman, \& Akmal}]{hanno_rein_2023_7778017}
Rein, H., Holman, M., \& Akmal, A. 2023, matthewholman/assist: v1.1.1, v1.1.1,
  Zenodo, \dodoi{10.5281/zenodo.7778017}

\bibitem[{{Rein} \& {Liu}(2012)}]{rebound}
{Rein}, H., \& {Liu}, S.~F. 2012, \aap, 537, A128,
  \dodoi{10.1051/0004-6361/201118085}

\bibitem[{{Rein} \& {Spiegel}(2015)}]{reboundias15}
{Rein}, H., \& {Spiegel}, D.~S. 2015, \mnras, 446, 1424,
  \dodoi{10.1093/mnras/stu2164}

\bibitem[{{Robbins} \& {Singer}(2021)}]{Robbins21}
{Robbins}, S.~J., \& {Singer}, K.~N. 2021, \psj, 2, 192,
  \dodoi{10.3847/PSJ/ac0e94}

\bibitem[{{Robinson} {et~al.}(2020){Robinson}, {Fraser}, {Fitzsimmons}, \&
  {Lacerda}}]{Robinson2020}
{Robinson}, J.~E., {Fraser}, W.~C., {Fitzsimmons}, A., \& {Lacerda}, P. 2020,
  \aap, 643, A55, \dodoi{10.1051/0004-6361/202037456}

\bibitem[{{Singer} {et~al.}(2019){Singer}, {McKinnon}, {Gladman},
  {Greenstreet}, {Bierhaus}, {Stern}, {Parker}, {Robbins}, {Schenk}, {Grundy},
  {Bray}, {Beyer}, {Binzel}, {Weaver}, {Young}, {Spencer}, {Kavelaars},
  {Moore}, {Zangari}, {Olkin}, {Lauer}, {Lisse}, {Ennico}, {New Horizons
  Geology}, Team, {New Horizons Surface Composition Science Theme Team}, \&
  {New Horizons Ralph and LORRI Teams}}]{Singer2019}
{Singer}, K.~N., {McKinnon}, W.~B., {Gladman}, B., {et~al.} 2019, Science, 363,
  955, \dodoi{10.1126/science.aap8628}

\bibitem[{{Skrutskie} {et~al.}(2006){Skrutskie}, {Cutri}, {Stiening},
  {Weinberg}, {Schneider}, {Carpenter}, {Beichman}, {Capps}, {Chester},
  {Elias}, {Huchra}, {Liebert}, {Lonsdale}, {Monet}, {Price}, {Seitzer},
  {Jarrett}, {Kirkpatrick}, {Gizis}, {Howard}, {Evans}, {Fowler}, {Fullmer},
  {Hurt}, {Light}, {Kopan}, {Marsh}, {McCallon}, {Tam}, {Van Dyk}, \&
  {Wheelock}}]{2MASS}
{Skrutskie}, M.~F., {Cutri}, R.~M., {Stiening}, R., {et~al.} 2006, \aj, 131,
  1163, \dodoi{10.1086/498708}

\bibitem[{{Smotherman} {et~al.}(2021){Smotherman}, {Connolly}, {Kalmbach},
  {Portillo}, {Bektesevic}, {Eggl}, {Juric}, {Moeyens}, \&
  {Whidden}}]{Smotherman2021}
{Smotherman}, H., {Connolly}, A.~J., {Kalmbach}, J.~B., {et~al.} 2021, \aj,
  162, 245, \dodoi{10.3847/1538-3881/ac22ff}

\bibitem[{{Smotherman} {et~al.}(2024){Smotherman}, {Bernardinelli}, {Portillo},
  {Connolly}, {Kalmbach}, {Stetzler}, {Juri{\'c}}, {Bekte{\v{s}}evi{\'c}},
  {Langford}, {Adams}, {Oldroyd}, {Holman}, {Chandler}, {Fuentes}, {Gerdes},
  {Lin}, {Markwardt}, {McNeill}, {Mommert}, {Napier}, {Payne}, {Ragozzine},
  {Rivkin}, {Schlichting}, {Sheppard}, {Strauss}, {Trilling}, \&
  {Trujillo}}]{Smotherman2024}
{Smotherman}, H., {Bernardinelli}, P.~H., {Portillo}, S. K.~N., {et~al.} 2024,
  \aj, 167, 136, \dodoi{10.3847/1538-3881/ad1524}

\bibitem[{{Stern} {et~al.}(2015){Stern}, {Bagenal}, {Ennico}, {Gladstone},
  {Grundy}, {McKinnon}, {Moore}, {Olkin}, {Spencer}, {Weaver}, {Young},
  {Andert}, {Andrews}, {Banks}, {Bauer}, {Bauman}, {Barnouin}, {Bedini},
  {Beisser}, {Beyer}, {Bhaskaran}, {Binzel}, {Birath}, {Bird}, {Bogan},
  {Bowman}, {Bray}, {Brozovic}, {Bryan}, {Buckley}, {Buie}, {Buratti},
  {Bushman}, {Calloway}, {Carcich}, {Cheng}, {Conard}, {Conrad}, {Cook},
  {Cruikshank}, {Custodio}, {Dalle Ore}, {Deboy}, {Dischner}, {Dumont},
  {Earle}, {Elliott}, {Ercol}, {Ernst}, {Finley}, {Flanigan}, {Fountain},
  {Freeze}, {Greathouse}, {Green}, {Guo}, {Hahn}, {Hamilton}, {Hamilton},
  {Hanley}, {Harch}, {Hart}, {Hersman}, {Hill}, {Hill}, {Hinson}, {Holdridge},
  {Horanyi}, {Howard}, {Howett}, {Jackman}, {Jacobson}, {Jennings}, {Kammer},
  {Kang}, {Kaufmann}, {Kollmann}, {Krimigis}, {Kusnierkiewicz}, {Lauer}, {Lee},
  {Lindstrom}, {Linscott}, {Lisse}, {Lunsford}, {Mallder}, {Martin}, {McComas},
  {McNutt}, {Mehoke}, {Mehoke}, {Melin}, {Mutchler}, {Nelson}, {Nimmo},
  {Nunez}, {Ocampo}, {Owen}, {Paetzold}, {Page}, {Parker}, {Parker},
  {Pelletier}, {Peterson}, {Pinkine}, {Piquette}, {Porter}, {Protopapa},
  {Redfern}, {Reitsema}, {Reuter}, {Roberts}, {Robbins}, {Rogers}, {Rose},
  {Runyon}, {Retherford}, {Ryschkewitsch}, {Schenk}, {Schindhelm}, {Sepan},
  {Showalter}, {Singer}, {Soluri}, {Stanbridge}, {Steffl}, {Strobel}, {Stryk},
  {Summers}, {Szalay}, {Tapley}, {Taylor}, {Taylor}, {Throop}, {Tsang},
  {Tyler}, {Umurhan}, {Verbiscer}, {Versteeg}, {Vincent}, {Webbert}, {Weidner},
  {Weigle}, {White}, {Whittenburg}, {Williams}, {Williams}, {Williams},
  {Woods}, {Zangari}, \& {Zirnstein}}]{Stern2015}
{Stern}, S.~A., {Bagenal}, F., {Ennico}, K., {et~al.} 2015, Science, 350,
  aad1815, \dodoi{10.1126/science.aad1815}

\bibitem[{{Stern} {et~al.}(2019){Stern}, {Weaver}, {Spencer}, {Olkin},
  {Gladstone}, {Grundy}, {Moore}, {Cruikshank}, {Elliott}, {McKinnon}, \&
  et~al.}]{Stern2019}
{Stern}, S.~A., {Weaver}, H.~A., {Spencer}, J.~R., {et~al.} 2019, Science, 364,
  aaw9771, \dodoi{10.1126/science.aaw9771}

\bibitem[{{Stern} {et~al.}(2023){Stern}, {White}, {Grundy}, {Keeney},
  {Hofgartner}, {Nesvorn{\'y}}, {McKinnon}, {Richardson}, {Marohnic},
  {Verbiscer}, {Benecchi}, {Schenk}, {Moore}, {New Horizons Geology}, \&
  {Geophysics Investigation Team}}]{Stern2023}
{Stern}, S.~A., {White}, O.~L., {Grundy}, W.~M., {et~al.} 2023, \psj, 4, 176,
  \dodoi{10.3847/PSJ/acf317}

\bibitem[{{Strauss} {et~al.}(2024){Strauss}, {Trilling}, {Bernardinelli},
  {Beach}, {Oldroyd}, {Sheppard}, {Schlichting}, {Gerdes}, {Fuentes}, {Holman},
  {Juri{\'c}}, {Lin}, {Markwardt}, {McNeill}, {Mommert}, {Napier}, {Payne},
  {Ragozzine}, {Rivkin}, {Smotherman}, {Trujillo}, {Adams}, \&
  {Chandler}}]{Strauss24}
{Strauss}, R., {Trilling}, D.~E., {Bernardinelli}, P.~H., {et~al.} 2024, \aj,
  167, 135, \dodoi{10.3847/1538-3881/ad1526}

\bibitem[{Team(2002--)}]{pytables}
Team, P.~D. 2002--, {PyTables}: Hierarchical Datasets in {Python}.
\newblock \url{https://www.pytables.org/}

\bibitem[{{Trilling} {et~al.}(2018){Trilling}, {Bannister}, {Fuentes},
  {Gerdes}, {Mommert}, {Schwamb}, \& {Trujillo}}]{Trilling18}
{Trilling}, D.~E., {Bannister}, M., {Fuentes}, C., {et~al.} 2018, arXiv
  e-prints, arXiv:1812.09705, \dodoi{10.48550/arXiv.1812.09705}

\bibitem[{Uieda {et~al.}(2020)Uieda, Soler, Rampin, van Kemenade, Turk,
  Shapero, Banihirwe, \& Leeman}]{uieda2020}
Uieda, L., Soler, S., Rampin, R., {et~al.} 2020, Journal of Open Source
  Software, 5, 1943, \dodoi{10.21105/joss.01943}

\bibitem[{{Verbiscer} {et~al.}(2019){Verbiscer}, {Porter}, {Benecchi},
  {Kavelaars}, {Weaver}, {Spencer}, {Buie}, {Tholen}, {Buratti}, {Helfenstein},
  {Parker}, {Olkin}, {Parker}, {Stern}, {Young}, {Ennico-Smith}, {Singer},
  {Cheng}, {Lisse}, \& {New Horizons Science Team}}]{Verbiscer2019}
{Verbiscer}, A.~J., {Porter}, S., {Benecchi}, S.~D., {et~al.} 2019, \aj, 158,
  123, \dodoi{10.3847/1538-3881/ab3211}

\bibitem[{{Verbiscer} {et~al.}(2022){Verbiscer}, {Helfenstein}, {Porter},
  {Benecchi}, {Kavelaars}, {Lauer}, {Peng}, {Protopapa}, {Spencer}, {Stern},
  {Weaver}, {Buie}, {Buratti}, {Olkin}, {Parker}, {Singer}, {Young}, \& {New
  Horizons Science Team}}]{Verbiscer2022}
{Verbiscer}, A.~J., {Helfenstein}, P., {Porter}, S.~B., {et~al.} 2022, \psj, 3,
  95, \dodoi{10.3847/PSJ/ac63a6}

\bibitem[{Virtanen {et~al.}(2020)Virtanen, Gommers, Oliphant, Haberland, Reddy,
  Cournapeau, Burovski, Peterson, Weckesser, Bright, {van der Walt}, Brett,
  Wilson, Millman, Mayorov, Nelson, Jones, Kern, Larson, Carey, Polat, Feng,
  Moore, {VanderPlas}, Laxalde, Perktold, Cimrman, Henriksen, Quintero, Harris,
  Archibald, Ribeiro, Pedregosa, {van Mulbregt}, \& {SciPy 1.0
  Contributors}}]{2020SciPy-NMeth}
Virtanen, P., Gommers, R., Oliphant, T.~E., {et~al.} 2020, Nature Methods, 17,
  261, \dodoi{10.1038/s41592-019-0686-2}

\bibitem[{{Weaver} {et~al.}(2022){Weaver}, {Porter}, {Spencer}, \& {The New
  Horizons Science Team}}]{2022PSJ.....3...46W}
{Weaver}, H.~A., {Porter}, S.~B., {Spencer}, J.~R., \& {The New Horizons
  Science Team}. 2022, \psj, 3, 46, \dodoi{10.3847/PSJ/ac4cb7}

\bibitem[{{Weaver} {et~al.}(2020){Weaver}, {Cheng}, {Morgan}, {Taylor},
  {Conard}, {Nunez}, {Rodgers}, {Lauer}, {Owen}, {Spencer}, {Barnouin},
  {Rivkin}, {Olkin}, {Stern}, {Young}, {Tapley}, \&
  {Vincent}}]{2020PASP..132c5003W}
{Weaver}, H.~A., {Cheng}, A.~F., {Morgan}, F., {et~al.} 2020, \pasp, 132,
  035003, \dodoi{10.1088/1538-3873/ab67ec}

\bibitem[{{W}es {M}c{K}inney(2010)}]{mckinney-proc-scipy-2010}
{W}es {M}c{K}inney. 2010, in {P}roceedings of the 9th {P}ython in {S}cience
  {C}onference, ed. {S}t{\'{e}}fan van~der {W}alt \& {J}arrod {M}illman, 56 --
  61, \dodoi{10.25080/Majora-92bf1922-00a}

\bibitem[{Wong {et~al.}(2017)Wong, Fan, Gao, Liang, Shia, Yung, Kammer,
  Summers, Gladstone, Young, Olkin, Ennico, Weaver, \& Stern}]{WONG2017}
Wong, M.~L., Fan, S., Gao, P., {et~al.} 2017, Icarus, 287, 110,
  \dodoi{https://doi.org/10.1016/j.icarus.2016.09.028}

\bibitem[{Yoachim {et~al.}(2022)Yoachim, Jones, Neilsen, Ribeiro, Daniel,
  Abrams, Almoubayyed, Andreoni, Awan, Becker, Bell, Bellm, Bianco, Bregeon,
  Bricman, Boberg, Carlin, Chen, Clarkson, Connolly, Gris, Hu, Kelley,
  Khakpash, Krughoff, Li, Lund, Marshall, Meyers, Prisinzano, Naghib, Rawls,
  Reuter, Rothchild, Setzer, Sick, \& Street}]{peter_yoachim_2022_7087823}
Yoachim, P., Jones, R.~L., Neilsen, E.~H., {et~al.} 2022, lsst/rubin\_sim:
  0.12.1, 0.12.1,  Zenodo, \dodoi{10.5281/zenodo.7087823}

\bibitem[{{Yoshida} {et~al.}(2024){Yoshida}, {Yanagisawa}, {Ito}, {Kurosaki},
  {Yoshikawa}, {Kamiya}, {Jiang}, {Stern}, {Fraser}, {Benecchi}, \&
  {Verbiscer}}]{Yoshida2024}
{Yoshida}, F., {Yanagisawa}, T., {Ito}, T., {et~al.} 2024, \pasj, 76, 720,
  \dodoi{10.1093/pasj/psae043}

\bibitem[{Zonca {et~al.}(2019)Zonca, Singer, Lenz, Reinecke, Rosset, Hivon, \&
  Gorski}]{Zonca2019}
Zonca, A., Singer, L., Lenz, D., {et~al.} 2019, Journal of Open Source
  Software, 4, 1298, \dodoi{10.21105/joss.01298}

\end{thebibliography}
\end{document}